\documentclass[sigconf,natbib=false,screen,authorversion]{acmart}

\AtBeginDocument{%
  }

    \usepackage{kantlipsum}
    \usepackage[capitalise,noabbrev]{cleveref} 
    \usepackage{csquotes} 
    \usepackage{subcaption} 
    \usepackage{wrapfig} 
    \usepackage{multirow} 
    \usepackage{float} 
    
    \usepackage{algorithm} 
    \usepackage{algpseudocode} 
    
\setcopyright{acmlicensed}
\copyrightyear{2024}
\acmYear{2024}
\acmDOI{XXXXXXX.XXXXXXX}

\acmConference[Conference acronym 'XX]{Make sure to enter the correct
  conference title from your rights confirmation emai}{June 03--05,
  2024}{Woodstock, NY}
\acmISBN{978-1-4503-XXXX-X/18/06}

\RequirePackage[
  datamodel=acmdatamodel,
  style=acmnumeric,
  natbib 
  ]{biblatex}

\begin{document}

\title{Slice it up: Unmasking User Identities in Smartwatch Health Data}

\author{Lucas Lange}
\affiliation{%
  \institution{Leipzig University \& ScaDS.AI Dresden/Leipzig}
  \city{Leipzig}
  \country{Germany}
}
\email{lange@informatik.uni-leipzig.de}
\orcid{0000-0002-6745-0845}

\author{Tobias Schreieder}
\affiliation{%
  \institution{Leipzig University \& ScaDS.AI Dresden/Leipzig}
  \city{Dresden}
  \country{Germany}
}
\email{tobias.schreieder@tu-dresden.de}
\orcid{0009-0000-8268-4204}

\author{Victor Christen}
\affiliation{%
  \institution{Leipzig University \& ScaDS.AI Dresden/Leipzig}
  \city{Leipzig}
  \country{Germany}
}
\email{christen@informatik.uni-leipzig.de}
\orcid{0000-0001-7175-7359}

\author{Erhard Rahm}
\affiliation{%
  \institution{Leipzig University \& ScaDS.AI Dresden/Leipzig}
  \city{Leipzig}
  \country{Germany}
}
\email{rahm@informatik.uni-leipzig.de}
\orcid{0000-0002-2665-1114}

%

\begin{abstract}
Wearables are widely used for health data collection due to their availability and advanced sensors, enabling smart health applications like stress detection. However, the sensitivity of personal health data raises significant privacy concerns. While user de-identification by removing direct identifiers such as names and addresses is commonly employed to protect privacy, the data itself can still be exploited to re-identify individuals. We introduce a novel framework for similarity-based Dynamic Time Warping (DTW) re-identification attacks on time series health data. Using the WESAD dataset  and two larger synthetic datasets, we demonstrate that even short segments of sensor data can achieve perfect re-identification with our Slicing-DTW-Attack. Our attack is independent of training data and computes similarity rankings in about 2 minutes for 10,000 subjects on a single CPU core. These findings highlight that de-identification alone is insufficient to protect privacy. As a defense, we show that adding random noise to the signals significantly reduces re-identification risk while only moderately affecting usability in stress detection tasks, offering a promising approach to balancing privacy and utility.
\end{abstract}

\ccsdesc[500]{Security and privacy~Usability in security and privacy}
\ccsdesc[500]{Human-centered computing~Ubiquitous and mobile devices}
\ccsdesc[500]{Security and privacy~Privacy protections}

\keywords{Privacy, User Re-identification, Dynamic Time Warping, Attack, Time Series, Similarity, De-identification}

\received{20 February 2007}
\received[revised]{12 March 2009}
\received[accepted]{5 June 2009}

\maketitle

\section{Introduction}\label{sec:intro}

The Internet of Things (IoT) is growing rapidly, finding widespread success in the area of wearable devices such as smartwatches.
Sales of smartwatches already amounted to 165 million units worldwide in 2023, while 209 million smartwatches are forecast to be sold in 2028~\cite{Statista2024SmartwatchMarket}.
Wrist-worn smartwatches are used for a variety of activities, including general productivity, but also tracking and recording personal health data, such as sleep, exercise, and stress.
For the purpose of health monitoring, smartwatches are equipped with various high-quality sensors that make it possible to record a person's sensitive health data over the long term.
While suitable devices are becoming more and more widespread and the amount of data collected is increasing rapidly as a result, the issue of data protection is becoming increasingly important and user awareness is growing.
Serious threats to such data are paramount, when looking at the various possible attacks on smart devices surveyed by \citet{Sikder2021SensorThreats}.
\citet{ernstInfluencePrivacyRisk2016a} also found that the perceived data protection risk has a direct influence on device acceptance and could ultimately prove to be a decisive factor for interested users.
In addition to general privacy concerns regarding personal data, there is also a direct correlation between perceived risk and trust in a data owner's privacy promise.
The primary reason for this may be the fact that as soon as a user's data is collected, the responsibility for privacy protection is completely transferred to the collecting institution, which is why the users should be fully informed about threats and defensive measures.

De-identification is currently the common mechanism for preserving privacy in such scenarios, with the aim of guarding a user's personal identity. For example, the privacy policy of a smartwatch distributor states that they \enquote{... may share non-personal information that is aggregated or de-identified so that it cannot reasonably be used to identify an individual}~\cite{fitbitprivacypolicy, fitbitresearchpledge}.
At first glance, this may conceal the identity through metadata removal, but it does not remove the inherent characteristics of an individual encoded into their health data.
For this reason, de-identification may not be an effective protection against identity inference~\cite{elemamSystematicReviewReIdentification2011}.

\begin{figure}[ht]
  \centering
  \includegraphics[width=0.42\linewidth]{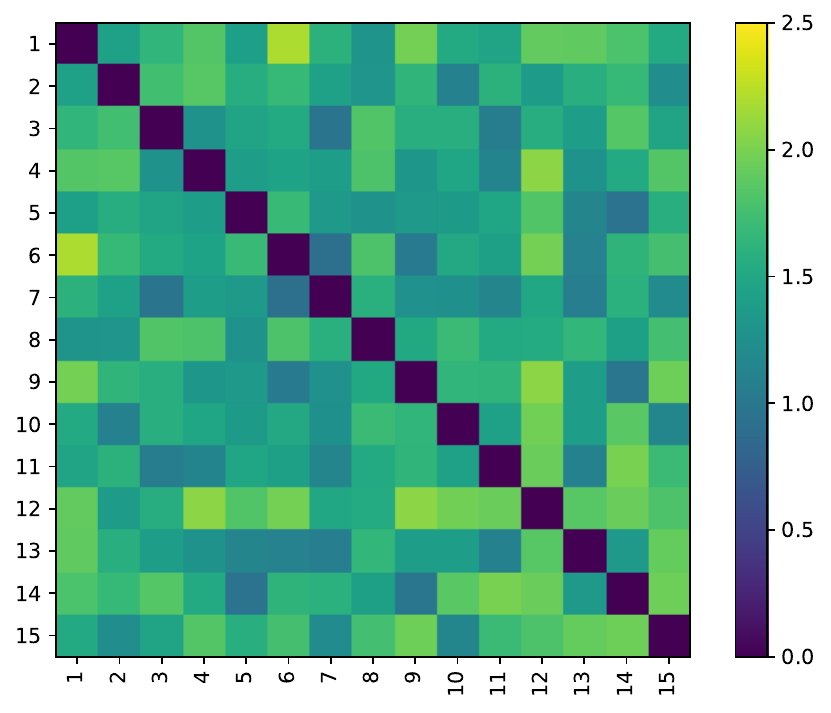}
  \caption{DTW distance heatmap for the 15 WESAD dataset subjects, where smaller scores indicate stronger similarity.}
  \label{fig:eda_dtw_alignment_heatmap_WESAD}
  \Description{Distances between WESAD subjects are diverse but never very low.}
\end{figure}

To demonstrate this risk, we apply  similarity-based re-identification attacks that solely use a short time series of a target user's health data to reconnect their de-identified data samples within a dataset.
Our attacks are based on the distance measure Dynamic Time Warping (DTW)~\cite{Meert2020dtaidistance} to compare the time series with each other and exploit the common characteristics of the provided multimodal sensor data.
\Cref{fig:eda_dtw_alignment_heatmap_WESAD} shows a matrix with calculated DTW distances between 15 subjects of the WESAD dataset~\cite{schmidtIntroducingWESADMultimodal2018}, where the diagonal line compares each subject's sample with itself.
Inside the matrix, distances between subjects vary constantly.
However, our proposed attacks show that even the often only small differences in distance, offer the potential to distinguish the original individuals.
We find our re-identification approach to be effective in breaking de-identification, especially in our example scenario in which institutions collect and leverage health data from smartwatches, which is why we emphasize stricter privacy measures.

\subsubsection*{Our contributions are:}
\begin{itemize}
    \item 
      We propose a framework for novel re-identification attacks based on DTW distances in time series health data. Unlike previous works, our approach does not rely on training data.
    \item
      We are the first to evaluate data-specific optimization strategies that exploit the multimodal and biological characteristics of the underlying health data.
    \item
      Our results expose inherent biometric re-identification threats in personal health data collected from smartwatches.
    \item 
      Our findings have practical relevance to smartwatch data collection scenarios, where user privacy is currently being implemented at scale through breakable de-identification.
\end{itemize}

In \cref{sec:background} we briefly review the relevant background before focusing on related work in \cref{sec:related}.
\Cref{sec:Attack Scenario} briefly describes our attack and introduces the example smartwatch scenario for our experiments.
These experiments and their outcomes are then outlined as our attack framework in \cref{sec:Overview of the Attack Framework,sec:evaluation}, respectively.
The following \cref{sec:discussion} is centered around discussing the implications of our results, which is divided into answering research questions and limitations regarding our approach.
Finally, we provide both a concise summary of our findings and an outlook into future work in the conclusive \cref{sec:conclusion}.

\section{Problem Statement}
    We first give fundamental notations and definitions for our attack.
    
    \begin{table}[ht]
        \centering
        \caption{Symbols and notations for our problem.}
        \label{tab:notations}
        \begin{tabular}{c|c}
        \hline
            Notation & Meaning  \\
          \hline \hline
            $M$ & set of available sensor modalities \\
          \hline 
            $A$    & known attack sample time series $A \in \mathcal{D}_A$  \\
          \hline
            $\mathcal{D}_A$  & dataset of known attack samples    \\
          \hline
            $T$ & de-identified target time series $T \in \mathcal{D}_T$    \\
          \hline 
            $\mathcal{D}_T$ & dataset of de-identified targets   \\
          \hline 
            $\mathcal{s}: A,T \rightarrow S$ & maps time series to their original subjects \\
          \hline 
            $T_A$ & correct target $T$ matching sample $A$\\& $T_A \iff \mathcal{s}(T)=\mathcal{s}(A)$   \\
          \hline
            $\mathcal{M}: \mathcal{D}_A \rightarrow \mathcal{D}_T$ & secret function that maps correct matches: \\ & $\mathcal{M}(A) = T_A$ or $\emptyset$ if $T_A \notin \mathcal{D}_T$ \\
           \hline
            $\mathcal{S}(A,\mathcal{D}_T)$ & similarity attack on $\mathcal{D}_T$ using $A$\\& returns similarity ranking $\mathbb{R}$ over $\forall T \in \mathcal{D}_T$ \\
        \hline
        \end{tabular}
    \end{table}
    
    \begin{definition}[Attack Sample and Target]\label{def:samples}
        Let a sequential time series of multimodal sensor data points be generated from a subject's ($s$) device, with $M$ giving the set of sensor modalities.
        Function $\mathcal{s}$ maps a sample to its originating subject.
        An attack sample is such a time series $A$ of subject $\mathcal{s}(A)$ collected by an adversary, who aims to match it to the correct target time series $T_A$ of the same subject $\mathcal{s}(A)$.
        All target time series $T$ stemming from subject $\mathcal{s}(T)$ are part of the de-identified dataset $\mathcal{D}_T$.
        Any $T \in \mathcal{D}_T$ is a target but may only be $T=T_A$ if $\mathcal{s}(T)=\mathcal{s}(A)$.
    \end{definition}
    
    \begin{definition}[Re-identification]\label{def:reidentification}
        Let $T$ be a target time series from a subject $\mathcal{s}(T)$.
        Then, $\mathcal{D}_T$ is the de-identified dataset of targets.
        Let $A$ be an attack sample time series from a subject $\mathcal{s}(A)$ known to the adversary.
        Then, $\mathcal{D}_A$ is the dataset of attack samples from known subjects. 
        Then the correct matching target $T_A$ for attack sample $A$ originally stemming from the same subject $s \in S$ is defined as: $T_A = T \in \mathcal{D}_T$ such that $\mathcal{s}(A)=\mathcal{s}(T)$. 
        There exists a secret matching function (unknown to the adversary) $\mathcal{M}: \mathcal{D}_A \rightarrow \mathcal{D}_T$ that maps known attack samples to their correct targets: $\mathcal{M}(A) = T_A$.
        $\mathcal{M}$ may be partial if $\exists A \in \mathcal{D}_A \forall T \in \mathcal{D}_T: \mathcal{s}(A) \neq \mathcal{s}(T)$, then $T_A \notin \mathcal{D}_T$ and $\mathcal{M}(A) = \emptyset$.
        The adversary does not know $\mathcal{M}$ but is successful at re-identification for subject $\mathcal{s}(A)$, when identifying the corresponding $T_A$ for their sample $A$ based on a similarity attack ranking: $\texttt{first}(\mathcal{S}(A,\mathcal{D}_T)) = \mathcal{M}(A)$.
    \end{definition}

\section{Background}\label{sec:background}
In this section we focus on introducing fundamental concepts.
    
\textbf{Dynamic Time Warping.}
    DTW~\cite{Giorgino2009dtw, Vintsyuk1968dtw} measures the similarity between temporal sequences by aligning them.
    It minimizes the differences between corresponding elements by accommodating temporal distortions through warping or stretching one sequence to better match another.
    This overcomes the limits of Euclidean distance, which solely compares data points at the same index~\cite{Eamonn2005dtw}.
     Although there are various other distance measures, Euclidean distance is the most common choice~\cite{Giorgino2009dtw}.
    DTW keeps distances between similar time series low, even when they have different lengths, distortions, or noise.
    
    To compute the distance between two time series $Q(q_{1}, q_{2}, \text{\ldots}, q_{n})$ and $C(c_{1}, c_{2}, \text{\ldots}, c_{m})$ of lengths n and m, an $n \times m$ matrix is constructed, containing distances between all data points in each series.
    The best alignment is then found using a warping path that minimizes the overall distance~\cite{Eamonn2005dtw}.
    Formally, DTW is represented as the minimization problem~\cite{Giorgino2009dtw, Eamonn2005dtw, tavenardblog2021dtw}:
        \begin{equation}
        	\text{DTW}(C, Q) = \min\limits_{\phi}(\sum\limits_{i, j \in \phi} d (c_{i}, q_{j})),
        	\label{eq:dtw}
        \end{equation}
        where the distance between $C$ and $Q$ is minimized along the warping path $\phi$, that sums up all the distances over $\phi$.
    
    In standard DTW, $n \times m$ distances must be calculated, resulting in time and space complexity of $O(n \times m)$ or more general $ O(n^2)$~\cite{Eamonn2005dtw}.
    However, through optimization strategies~\cite{Rakthanmanon2021dtw}, the amortized costs of DTW can be reduced to an average complexity of less than $O(n)$.
    For our DTW$_{\theta}$ variants on multimodal time series, we add a dimension over the set of signal modalities $M$ (\cref{sec:dtw_attacks}).

\textbf{Generative Adversarial Networks.}
    Abundant training data is essential for machine learning, but in health data, creating such datasets is challenging due to limited subjects, ethical concerns, financial constraints, and privacy issues~\cite{Imtiaz2021GAN}.
    Generative Adversarial Networks (GANs)~\cite{Goodfellow2014GAN} offer a solution through synthetic data.
    
    GANs consist of two neural networks: a generator ($G$), which creates synthetic samples from random noise, and a discriminator ($D$), which classifies samples as real or synthetic.
    These networks are trained adversarially, with $G$ improving its outputs to deceive $D$, and $D$ refining its ability to distinguish real from synthetic data.
    
    Extensions of the original GAN architecture, such as conditional GANs (CGAN)\cite{Mirza2014cGAN} and DoppelGANger (DGAN)\cite{Lin2020DGAN}, are also able to reflect the class structure of the original dataset.

\textbf{De-identification and Identity Inference.}
    De-identification is an anonymization technique that removes direct identifiers such as names, locations, or other metadata to protect individuals when sensitive data is collected or released publicly.
    However, re-identification still frequently occurs, especially with health data~\cite{elemamSystematicReviewReIdentification2011}.
    In attacks categorized as identity inference, adversaries infer identities or link records to specific individuals, undermining the de-identification.
    \citet{HenriksenBulmer2016ReIdentification} reviewed such attacks from 2009 to 2016, finding that 72.7\% were successful, highlighting the need for improved mitigation strategies.
    Membership inference~\cite{shokri2017membership} as a related attack paradigm, only aims to determine whether a target is present in a dataset, rather than identifying specific individuals.

\section{Related Work}\label{sec:related}
The increasing use of wearable devices leads to the easy generation and sharing of data collected from individuals. \citet{Chikwetu2023survey} reviewed 72 studies about re-identification methods based on collected data from wearable devices to estimate the risk that an individual is re-identified within a data collection. The risk estimation allows us to derive potential consequences regarding data-sharing policies in the context of data privacy and FAIR principles. The study observed that most methods require very little data to re-identify an individual highlighting the risk of revealing information. 

In the biomedical domain, various methods~\cite{Randazzo2020ECG, Zhang2019EEG} utilize electroencephalogram (EEG) and electrocardiogram (ECG) data. The work of \citet{Zhang2019EEG} builds a certain classification model for each individual based on manually defined features. \citet{Randazzo2020ECG} train a neural network based on ECG-derived data where each individual or similar ones represents one class. Due to the requirement of individual classification models, these approaches are not feasible for a large amount of data.

In addition to biomedical signals, accelerometry data, and gyroscope data can be used for re-identifying individuals~\cite{Saleheen2021ReIdentification, Weiss2019smartwatch} or to predict the location of metro riders~\cite{Hua2017tracking}. The work of \citet{Saleheen2021ReIdentification} uses accelerometry data from 353 participants being recorded for 190,078 hours (70 days with at least 8 hours per day) resulting in 51.3 billion data points. The attack aims to determine the trace from an anonymized database regarding an available trace where the user is known. The approach computes similarities between the anonymized and known time series. Therefore, the traces are split into smaller segments to build meaningful features using a neural network. The network consists of convolutional layers and gated recurrent units to address the time aspect. Moreover, the base model classifies resulting features if the segment from the known user corresponds to the anonymized one. The authors suggest various aggregation strategies to determine the similarity between traces based on the segment similarities.

In contrast to this approach, we do not utilize a supervised feature extraction and classification model where the performance depends on training data, which is rarely available. Our proposed method can be evaluated for each available individual because we do not split the data into training and test datasets. Moreover, we consider various sensor data types and thus not only focus on accelerometry data. In contrast to our evaluation, the work only considers the true matching rate which restricts the attacker from a more differentiated view though taking the top k results.


In the domain of authentication, biometric images such as fingerprints are used as keys to log in to systems or applications. The original images are encoded to templates using, e.g., Bloom filter, neural networks, etc.~\cite{Sandhya2017biometrics}. 
Due to the preservation of the similarity between original images and templates, similarity-based attacks aim to construct an image where the encoded template is similar to the target template. Therefore, similarity-based attack methods~\cite{Dong2019biometrics, Yang2023palmprint} compare a fake template with a target template and iteratively optimize the construction process to obtain a new image for generating a new template.

DTW has been applied in re-identification attacks for side channels~\cite{zhang2018level} and trajectory data~\cite{avola2024signal,brauer2024time,lestyan2019extracting} but while they compare full samples, we take slices of available signals to improve alignment.

In summary, most of related work formulates re-identification as a multi-class problem where each individual is characterized by a certain class~\cite{Randazzo2020ECG, Weiss2019smartwatch} or as an authentication problem where each individual is represented by its model~\cite{Weiss2019smartwatch, Zhang2019EEG}. Moreover, the approaches require a feature engineering step based on manually defined features~\cite{Randazzo2020ECG, Weiss2019smartwatch, Zhang2019EEG} or supervised learned ones~\cite{Saleheen2021ReIdentification}.



\section{Attack Scenario}
\label{sec:Attack Scenario}

    \begin{figure}[ht]
      \centering
      \includegraphics[width=1.0\linewidth]{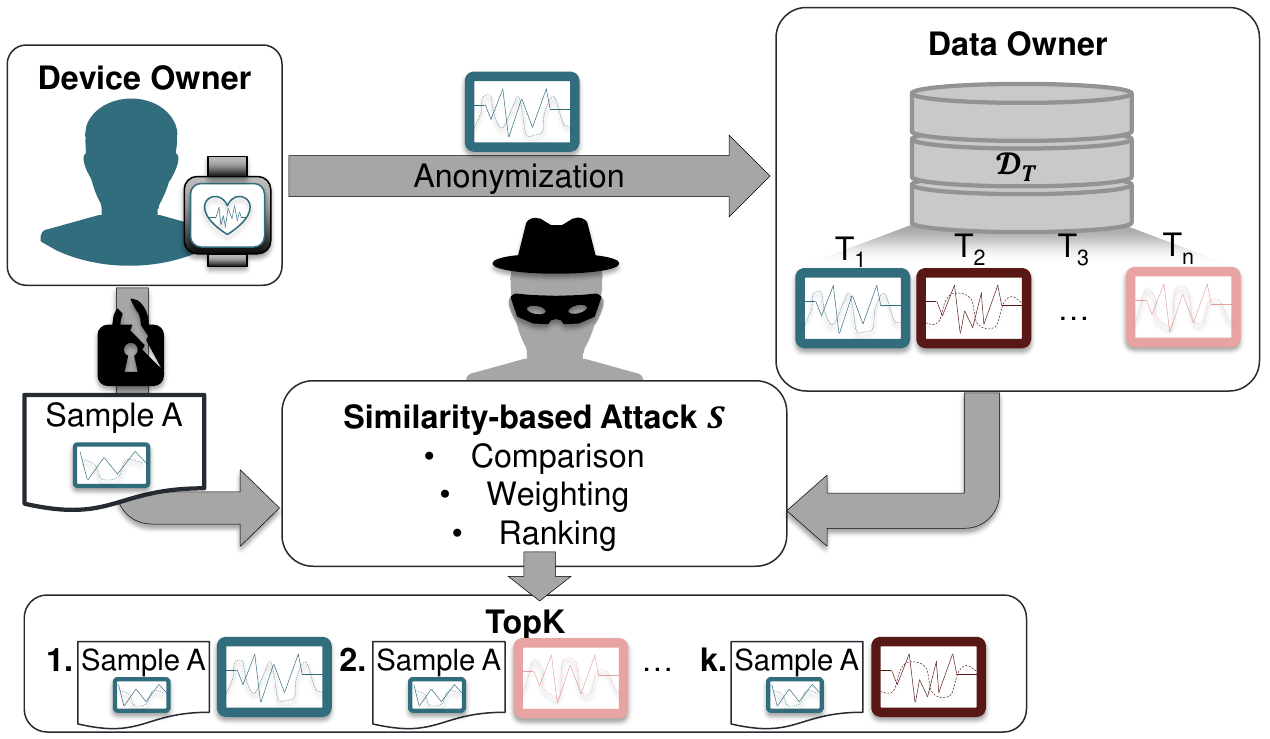}
      \caption{The attack scenario consists of a device owner, a data owner, and an attacker. The device owner sends time series data via his device to the data owner, who anonymizes and stores it in a dataset for analysis. The attacker aims to retrieve the device owner's data from this dataset utilizing a known data sample of the device owner.}
      \label{fig:scenario}
      \Description{An attacker is in the middle of this setting. He gets the sample from a device owner and has access to the data owner's database, where device owner data is stored. He then performs a similarity attack and ranks the database records to find the correct match between device owner sample and database entries of the same device owner.}
    \end{figure}

    In our scenario, the attacker uses their attack data samples $A \in \mathcal{D}_A$ with anonymized records within the dataset $\mathcal{D}_T$. The attacker performs a similarity-based attack $\mathcal{S}(A, \mathcal{D}_T)$, ranking individuals by their DTW distances to re-identify the correct target time series $T_A$, thus undermining data privacy promises. Figure~\ref{fig:scenario} illustrates the attack scenario involving three actors:

    \textbf{Device Owner.} An individual using a smartwatch to record health data, which is transferred to apps and cloud storage.
    
    \textbf{Data Owner.} An entity separate from the device owner collecting data from smart devices and using it to enhance services. The data is anonymized by de-identification to ensure privacy.
    
    \textbf{Attacker.} An adversary with a data sample of the device owner, aiming to break de-identification. This might be an insider on the data owner side or the data owner themselves, trying to break their privacy promise. Another possibility would be a third party seeking out sensitive information about the device owner for personal gain.

    The device owner is the targeted user in our attack.
    They generate health data on a smartwatch that is forwarded to the data owner in de-identified form, i.e., with all directly identifying information removed and replaced by unique IDs.
    However, the data itself is passed on unchanged to keep the usability for analysis by the data owner.
    The data owner stores this and other subjects' data in an aggregated dataset $D_T$.
    Our DTW-based attacks enable an attacker to find the original identity of such target data ($T$) in the dataset ($D_T$) containing all device owners as subjects.
    The only prerequisite is access to a (short) attack data sample ($A$) of the device owner.
    The attacker might as well possess a set of attack samples ($\mathcal{D}_A$) targeting multiple individuals.
    The attacker then carries out a similarity-based attack $\mathcal{S}(A,\mathcal{D}_T)$ by calculating the DTW distances between an attack sample $A \in \mathcal{D}_A$ and the anonymized targets $T \in \mathcal{D}_T$. 
    By performing a ranking on these distances, they may identify the target $T$ having the highest similarity, i.e., lowest distance to $A$.
    This target is the most likely to be the device owner, i.e., the correct target $T_A$.
    If correctly re-identified, any promised data privacy guarantees are negated and user privacy would be broken irreparably, rendering the anonymization of the device owner useless.
    It does not matter if data is stored or just processed, since the attack can also be executed directly on any arriving data samples.
    
    While an insider on the device owner's side would have the easiest access to an attack sample, there are various other scenarios in which the attacker could gain access to the device owner's sample:

    \textbf{Self-publication by device owner.} The device owner (accidentally) publishes the data. This could be through a fitness app synced with the smartwatch~\cite{Saleheen2021ReIdentification}. The owner may publicly share data via the app, making it available to potential attackers.
    
    \textbf{Leak by data owner.} Smart devices offer cloud storage for accessing data across devices. However, the security of these services depends on the data owner and might not be communicated to users. Cloud data can be vulnerable backdoor exploits, as well as, insecure authentication systems leading to data breaches~\cite{Ching2016SmartwatchAttack}.
    
    \textbf{Security vulnerabilities.} Software or hardware vulnerabilities in smart devices or apps are an issue due to the lack of security features, such as secure authentication, PINs, and data encryption~\cite{Ching2016SmartwatchAttack, Liu2016SmartwatchAttack, Blow2020SmartwatchAttack}. If an attacker steals or replaces a device, they may access on-device data. Additionally, the Bluetooth connection for data transfer to smartphones could be intercepted through man-in-the-middle attacks using sniffers~\cite{Blow2020SmartwatchAttack, Ching2016SmartwatchAttack, Silva-Trujillo2023SmartwatchAttack}. Malware and phishing attacks targeting can also lead to unauthorized data acquisition.
    
\section{Attack Framework}\label{sec:Overview of the Attack Framework}

\begin{figure*}[ht]
  \centering
  \includegraphics[width=0.8\linewidth]{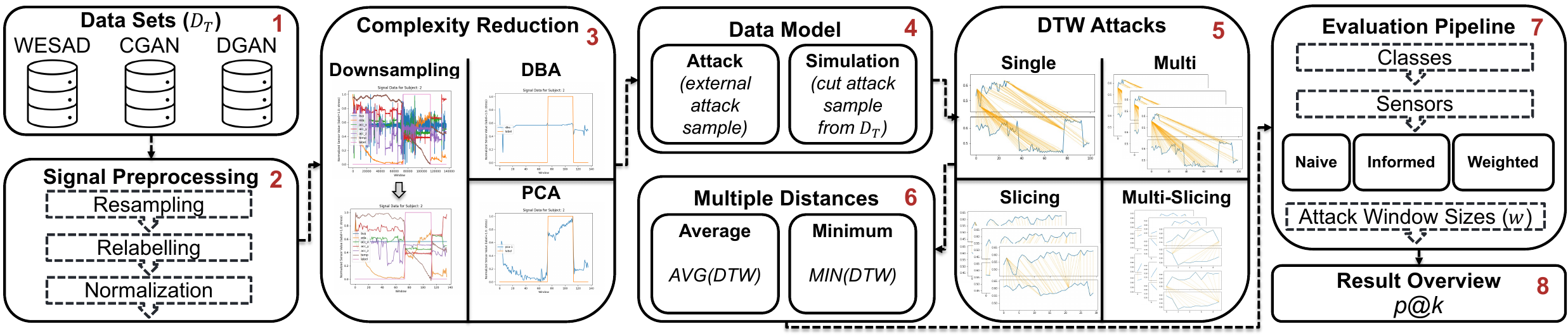}
  \caption{Overview of the evaluated attack framework.}
  \label{fig:framework-overview}
  \Description{The eight-step process is described in the main text.}
\end{figure*}
    
    \Cref{fig:framework-overview} illustrates the eight-stage attack framework for our experiments, where we favored a modular design for allowing the integration of datasets and methods using interfaces: 
    (1) Select a target dataset $\mathcal{D}_T$ for the re-identification attack.
    (2) Preprocess the dataset to ensure data consistency.
    (3) Apply optional complexity reduction.
    (4) If no external attack sample is provided, the data model takes care of creating a $\mathcal{D}_A$ simulation, cutting attack data out of $\mathcal{D}_T$.
    (5) We offer different DTW attack ($\mathcal{S}$) strategies based on DTW$_{\theta}$ variants for calculating the distance scores.
    (6) If DTW$_{\theta}$ produces multiple results, we offer aggregation methods.
    (7) The attack evaluation uses a three-stage rank-based evaluation pipeline for incorporation classes, sensor modalities, and attack window sizes.
    (8) A result overview for the various evaluation results.
    
    \subsection{Datasets}\label{sec:datasets}
    We employ two kinds of data in our experiments, with the first being real lab data from the relatively small WESAD dataset. The second type is synthetic data generated from training GANs on the WESAD data for testing our attacks on larger synthetic datasets.
    
        \textbf{WESAD.}\label{sec:wesad}
            The WESAD dataset (WEarable Stress and Affect Detection) was introduced by \textcite{schmidtIntroducingWESADMultimodal2018} and has since been widely used for stress detection research~\cite{Souza2022StressDetection, Eren2022StressDetection, GilMartin2022HumanStress, lange2023transformer, Li2020StressDetection, Siirtola2019StressDetection}.
            It includes multimodal data from two wearable sources: a chest-worn RespiBan and a wrist-worn Empatica E4, which both track divers signal modalities at varying sampling rates.
            
            We utilize only Empatica E4 data, targeting privacy risks in smartwatches. The watch collects blood volume pulse (BVP, 64Hz), electrodermal activity (EDA, 4Hz), skin temperature (TEMP, 4Hz), and 3-axis accelerometer (ACC[x,y,z], 32Hz) and the dataset includes recordings from 15 subjects (12 males, 3 females; mean age: 27.5) across 36 minutes each.
            The study had five phases covering different affective states: \textit{baseline} (20 min of neutral state), \textit{amusement} (6.5 min of humorous video clips), \textit{stress} (Trier Social Stress Test), \textit{meditation} (7 min of guided breathing), and \textit{recovery} (final debriefing), which are commonly simplified to \textit{stress} and \textit{non-stress}.
        
        \begin{figure}[ht]
            \centering
            \begin{subfigure}{0.42\linewidth}
                \includegraphics[width=\linewidth]{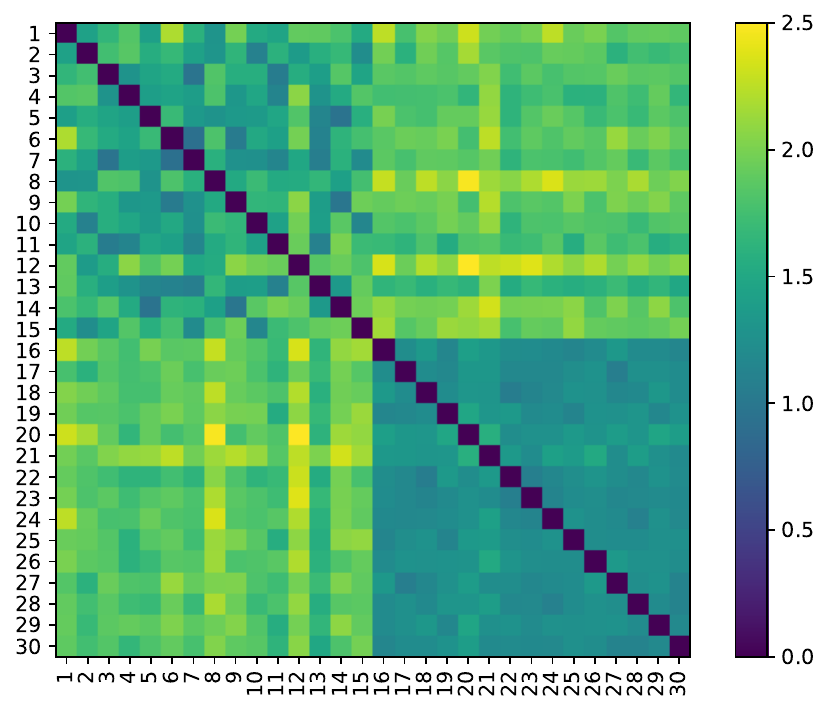}
                \caption{WESAD-CGAN$_{15}$}\label{fig:heatmap_cGAN}
            \end{subfigure}
            \begin{subfigure}{0.42\linewidth}
                \includegraphics[width=\linewidth]{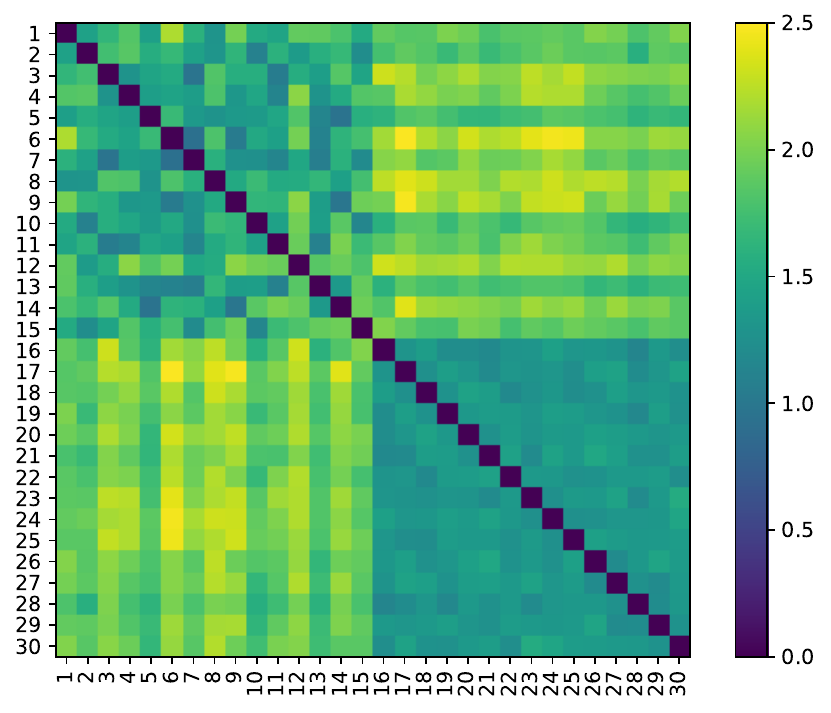}
                \caption{WESAD-DGAN$_{15}$}\label{fig:heatmap_dGAN}
            \end{subfigure}
            \caption{DTW distance heatmaps between the subjects of the WESAD dataset (1-15) and the CGAN$_{15}$ or DGAN$_{15}$ dataset (16-30). Distance values are averaged over sensors and the same subjects are matched on the diagonal.}
            \label{fig:dtw_heatmap_gan}
            \Description{The differences between DTW distances depending on the dataset are conveyed in the main text.}
        \end{figure}
        \textbf{Synthetic.}
            %
            %
            %
            Since the 15 subjects in the WESAD dataset are a rather small testing sample and without other sources publicly available, we want to approximate our scalability regarding attack success and runtime using generated data.
            For creating synthetic data, we leverage existing work~\cite{Lange2024GANStressDetection}, which employed CGAN and DGAN models to augment the WESAD dataset with comparable synthetic subject data.
            We denote synthetic datasets using the format GAN$_{\text{\#subjects}}$, where the number indicates the dataset size.
            For example, a dataset with 15 synthetic subjects generated by the CGAN is referred to as CGAN$_{15}$.
            We always include the same subjects for the same size, and larger datasets are created by adding subjects in the same fixed order.
            The authors~\cite{Lange2024GANStressDetection} ensured their synthetic data mirrors the original properties, maintaining similar signal correlations and overall structure.
            When we compare DTW distance variations between subjects in the synthetic datasets and the original WESAD data, ideally, the distance distribution across subjects should be consistent between datasets.
            However, as shown in \cref{fig:dtw_heatmap_gan}, when comparing 15 real WESAD subjects to 15 synthetic subjects from CGAN and DGAN, we observe that the synthetic data exhibits significantly less variation in DTW distances.
            The WESAD dataset, shown at the top left of the matrices, displays substantial variation in distances, whereas the GAN datasets at the bottom right show uniformly low distances among subjects.
            
            When comparing WESAD data to the synthetic datasets (seen in the lower left and top right corners of the matrices), distances are high for both, with CGAN showing slightly lower overall scores. While this might seem like a drawback, it can actually be advantageous.
            The higher distances from the original data introduce desired diversity, and the uniformly low distances within the CGAN and DGAN datasets create a more challenging re-identification task compared to the WESAD dataset, where subjects are less similar.
            Therefore, we use CGAN and DGAN data to test the scalability of our approach to larger subject numbers, expecting a slightly lower success rate than with the original data, which might provide a conservative lower bound when compared to real data.
    
    \subsection{Signal Preprocessing}\label{sec:Signal Preprocessing}
        For the preprocessing stage (2), we largely adopt the signal preprocessing of \textcite{GilMartin2022HumanStress}, which consists of three steps: signal resampling, relabelling, and normalization. As described in \cref{sec:wesad}, the sensors of the Empatica E4 track signals at different sampling rates. We therefore resample them to a consistent 64 Hz sampling rate by applying a fast Fourier transform, which ensures a time series with exactly one data point per signal at each point in time.
        In the second step, we adjust the labels of the resampled signal data. The original labels available in the WESAD dataset are divided into \textit{baseline}, \textit{amusement}, \textit{stress}, \textit{meditation} and \textit{recovery}.
        For supporting a binary stress detection task, we want to consolidate the labels into \textit{stress} and \textit{non-stress}.
        We thus initially drop the very few data points labelled \textit{meditation} and \textit{recovery}, and combine the labels \textit{baseline} and \textit{amusement} into \textit{non-stress}.
        This leads to an average of 70\% \textit{non-stress} and 30\% \textit{stress} data per subject.
        Third, we perform a min-max normalization in the [0, 1] range to remove the scaling difference between signal modalities, which speeds up calculations, as well as, increasing the signal comparability of different subjects in our similarity ranking.
        Our preprocessing is adapted to our expected data in the targeted database but can be changed to fit different data sampling and labels.
    
    \subsection{Complexity Reduction}\label{sec:Complexity Reduction}
    \begin{figure}[ht]
      \centering
      \includegraphics[width=0.9\linewidth]{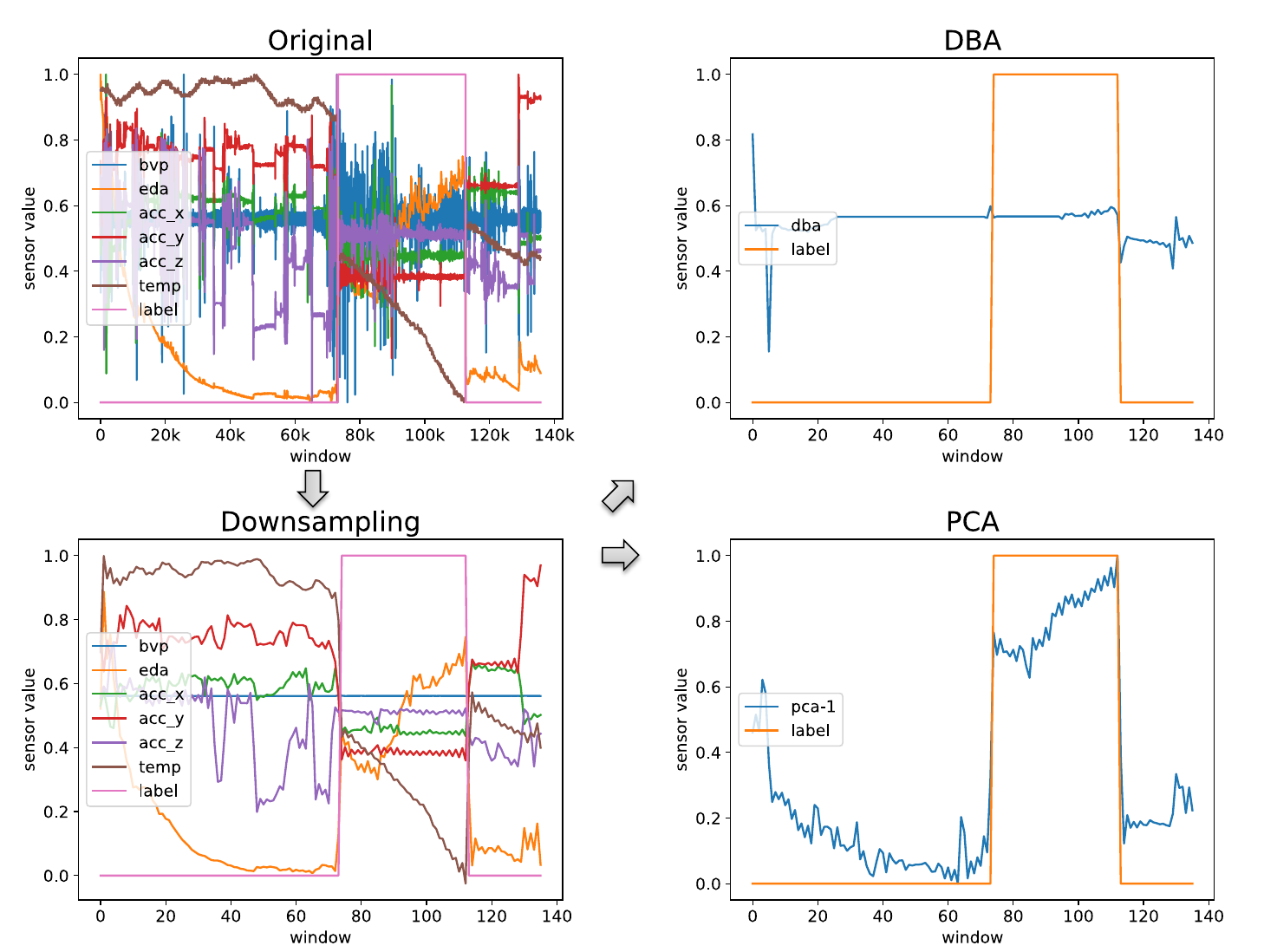}
      \caption{Example results of our complexity reduction methods on subject s$_2$ of the WESAD dataset.}
      \label{fig:complexity_reduction}
      \Description{The differences between the methods are conveyed in the main text.}
    \end{figure}
    
        For large databases we find notable performance pre-requisites for our DTW attacks, which leads us to the integration of complexity reduction methods, as shown in stage (3) of \cref{fig:framework-overview}. We discuss the methods of downsampling, DBA, and PCA in detail below. Their results are exemplified in \cref{fig:complexity_reduction}, where we apply DBA and PCA after downsampling the signals for reducing noise.
            
        \textbf{Downsampling.}
            Downsampling is our baseline step for complexity reduction and reduces the sampling rate of our signals by a set downsampling factor (DSF) to shorten the signals. For example, a downsample factor of $DSF=1000$ reduces 136,000 to 136 data points. This enhances runtimes and smooths signals, reducing noise while retaining the original course, as seen in \cref{fig:complexity_reduction}.
            
        \textbf{Dynamic Time Warping Barycenter Averaging.}
            DBA~\cite{Petitjean2011DBA} averages sequences over their multimodal signals to reduce the overhead from multiple modalities and thereby improves our runtimes. It uses an heuristic strategy that iteratively refines an initial arbitrary average sequence to minimize the sum of squared DTW distances from the average sequence to the overall set of sequences. When applied on our downsampled signals in \cref{fig:complexity_reduction}, we can notice how DBA focuses on just a few deflections in the signals.
            
        \textbf{Principal Component Analysis.}
            PCA~\cite{wold1987principal} projects data into lower dimensions, while retaining its significant variance. Using the first PC, we simplify our data representation, facilitating faster computations. \Cref{fig:complexity_reduction} clearly shows the differences between PCA and DBA, with substantial changes in PC amplitudes between labels.
            
    \subsection{Data Model}\label{sec:data model}
        The correct data model in stage (4) is a key factor in simulating our attack scenario and has two different modes, the \textit{external} and the \textit{simulation} mode. These two modes determine which data is used as the attack sample for the re-identification attacks.

        \textbf{External.}
        The external mode offers an realistic in-practice attack, where the attacker owns an attack sample $A$ that is used as external input for the attacks.
        The method is described in \cref{alg:external} and may be repeated for multiple $A \in \mathcal{D}_A$ if the attacker posses a set.

    \begin{figure}[t]
        \vspace{-1.3em} 
        \begin{minipage}{\linewidth}
           \begin{algorithm}[H]
            \caption{External Mode (external $A$ input)}\label{alg:external}
            \begin{algorithmic}[1]
                \State \textbf{Input:} $\mathcal{D}_T$: Set of $n$ targets $T$, $A$: Attack sample
                \State \textbf{Output:} $T$: Most similar target $T \in \mathcal{D}_T$ to $A$
                
                \State $\mathcal{D}_T,$ $A := $ \texttt{Preprocess}($\mathcal{D}_T, A$)\Comment{\cref{sec:Signal Preprocessing}}
                \State $\mathcal{D}_T,$ $A := $ \texttt{Complexity}($\mathcal{D}_T, A$)\Comment{Optional, \cref{sec:Complexity Reduction}}
                
                \State $\mathbb{R} = \mathcal{S}(A, \mathcal{D'}_T)$\Comment{\cref{sec:dtw_attacks}}
                \State $T\gets \texttt{first}(\mathbb{R})$\Comment{First ranked target}
                
                \State \Return T
                    \hspace{\algorithmicindent} \Comment{Successful re-identification if $T=T_A$}
            \end{algorithmic}
           \end{algorithm}
           
           \vspace{-2em} 
            
           \begin{algorithm}[H]
            \caption{Simulation Mode (no $A$ input)}\label{alg:simulation}
            \begin{algorithmic}[1]
                \State \textbf{Input:} $\mathcal{D}_T$: Set of $n$ targets $T$, $T$: Attacked target $T \in \mathcal{D}_T$
                \State \textbf{Output:} P: Set of p@k scores regarding $T_A$ ranking
                
                \State $\mathcal{D}_T := $ \texttt{Preprocess}($\mathcal{D}_T$)\Comment{\cref{sec:Signal Preprocessing}}
                \State $\mathcal{D}_T := $ \texttt{Complexity}($\mathcal{D}_T$)\Comment{Optional, \cref{sec:Complexity Reduction}}
                
                \State $\mathcal{D'}_T,$ $A_{non}$, $A_{stress} = $  \texttt{Cut}($\mathcal{D}_T$, $T$, $w$, $w_{\text{adj}}$)\Comment{\cref{sec:data model}}
                \State $\mathbb{R}_{non} = \mathcal{S}(A_{non},\mathcal{D'}_T)$\Comment{\cref{sec:dtw_attacks}}
                \State $\mathbb{R}_{stress} = \mathcal{S}(A_{stress},\mathcal{D'}_T)$

                \State $T_A\gets \mathcal{M}(A_{non})$\Comment{Correct target for $A_{non}$/$A_{stress}$}
                
                \State $r_{non}\gets \texttt{getRank}(\mathbb{R}_{non},T_A)$ \Comment{Rank of $T_A$ in $\mathbb{R}$}
                \State $r_{stress}\gets \texttt{getRank}(\mathbb{R}_{stress},T_A)$
                
                \State $P = \mathcal{W}_{classes}(n,r_{non},r_{stress})$  \Comment{Get p@k scores, \cref{sec:Classes}}
                
                \State \Return $P$\Comment{Successful re-identification if $p@1=1$}
             \end{algorithmic}
            \end{algorithm}
        \end{minipage}
    \end{figure}

    \begin{algorithm}[t]
        \caption{(Threshold) DTW Re-identification Attack}\label{alg:attack}
        \begin{algorithmic}[1]
            \State \textbf{Input:} $\mathcal{D}_T$: Set of $n$ targets $T$, $A$: Attack sample, $\mathscr{t}$: Threshold
            \State \textbf{Output:} $\mathbb{R}$: Distance-based ranking or $\emptyset$
            
            \Function{$\mathcal{S}$}{$A,\mathcal{D}_T,\mathscr{t}$}\Comment{For standard functionality $\mathscr{t}=\infty$}
                \State $R\gets \emptyset$\Comment{Initialize}
                \For{$i=1$ \textbf{to} $n$}
                    \State $D\gets \text{DTW}_{\theta}(A, T_i)$\Comment{Calculate distances, \cref{sec:dtw_attacks}}
                    \State $d\gets \texttt{Agg}(D, W)$\Comment{Aggregate over signals, \cref{sec:Sensor Ranking}}
                    \If{$d \leq \mathscr{t}$}\Comment{Test distance threshold, \cref{sec:in-out}}
                        \State $R\texttt{.append}((T_i:d))$\Comment{$T$-Distance pairs, \cref{sec:ranking}}
                    \EndIf
                \EndFor
                \If{$R = \emptyset$}\Comment{No similar target for threshold}
                    \State $\mathbb{R}\gets \emptyset$\Comment{No target found, \cref{def:reidentification}}
                \Else
                    \State $\mathbb{R}\gets $ \texttt{Rank}(\texttt{sort}($R$))\Comment{Rank targets $T$ on $d$, \cref{sec:ranking}}
                \EndIf
                \State \Return $\mathbb{R}$
            \EndFunction
        \end{algorithmic}
    \end{algorithm}

        \textbf{Simulation.}
        By contrast, in simulation mode, the attack sample $A$ is cut out of the given target dataset $\mathcal{D}_T$ itself.
        This mode is needed for determining the performance of our DTW attacks in experiments, since we have to know the correct target $T_A$ for each sample.
        The simulation abides to three key parameters for cutting attack samples from the original data.
        The first is the \textit{attack window size} $w$, which reflects the length of an attack sample $A$, i.e., the number of windows cut out of the target's signal.
        The second parameter is called \textit{adjacent windows} $w_{\text{adj}}$ and determines how many additional adjacent windows are cut from the signals at the edges of the sample, which prevents accidental alignment due to adjacent similar data points.
        By default, we use $w_{\text{adj}}=1000$ adjacent windows for the experiments, which changes in relation to downsampling.
        With a $DSF=1000$, we just remove one adjacent window ($w_{\text{adj}}=1$).
        The \textit{classes} (stress or non-stress) constitute our third parameter.
        For a realistic result and evaluating the possibly differing threat levels, we always attack using both stress and non-stress data---further explained in \cref{sec:Classes}. 
        
        As detailed in \cref{alg:simulation}, we first select the target $T \in \mathcal{D}_T$ as the attacked subject.
        When creating an attack sample $A$ for a target $T \in \mathcal{D}_T$ (i.e., a subject), we first take the time series $T$ and split it into its non-stress and stress segments.
        We then cut out a sample of length $w$ from the middle of each segment, while also cutting the additional adjacent windows $w_{\text{adj}}$ that are thrown away.
        We now have $A_{non}$ and $A_{stress}$ for $T$, and the rest of the cut-up time series is simply merged at the new borders, again creating one continuous time series $T'$.
        This can lead to jumps in the signals, which, however, already commonly exist in the original data due to measurement errors and dropping the meditation and recovery labels in our preprocessing.
        In fact, random signal jumps potentially make similarity-based re-identification more difficult.
        
        To ensure that our attack sample creation has no influence on our re-identification results, we also cut samples out of all other $T \in \mathcal{D}_T$ using the same process.
        We thereby remove the potential influence of our cuts and especially the signal lengths on the DTW distances.
        Otherwise, the remaining data for $T_A$ would be shorter than other target data, which leads to an underestimation of DTW distance solely based on less data points comparisons.
        
        Finally, we evaluate the resulting similarity rankings for $A_{non}$ and $A_{stress}$ regarding the targets $T \in \mathcal{D}_T$ after our attack by determining p@k scores as detailed in \cref{sec:ranking,sec:Classes}.
    
    \subsection{DTW-Attacks}\label{sec:dtw_attacks}
    
        This section presents our four DTW-based re-identification attacks 
            that an adversary may use for comparing an attack sample $A$ to entries $T$ in a target dataset $\mathcal{D}_T$ regarding their similarity.
        The overall attack follows \cref{alg:attack} in its standard variation for given inputs $A$ and $\mathcal{D}_T$.
        Here, we focus on the computation of DTW distances using custom DTW$_{\theta}$ variants as in Line 6 of \cref{alg:attack}.
        
        \textbf{Single-DTW-Attack.}\label{sec:Single-DTW-Attack}
            This is our baseline that calculates standard DTW alignments over a set of signal modalities $M$ between an attack sample $A$ and an entry $T \in \mathcal{D}_T$ using \cref{eq:dtw}:
            \begin{equation} 
                \text{DTW}_{\text{Single}}(A, T) = \{\text{DTW}(A_m, T_m) | \forall m \in M\},
                \label{eq:dtw_single}
            \end{equation}
            
            Since we have multimodal time series, we create individual distance scores between the same modalities $m \in M$.
            This results in a $n \times m$ distance matrix, with $n=|\mathcal{D}_T|$ and $m=|M|$, where smaller distances indicate higher similarity.
        
        \textbf{Multi-DTW-Attack.}\label{sec:Multi_DTW_Attack}
            The \textit{Multi-DTW-Attack} divides the attack sample $A$ into multiple subsets of equal length based on a predefined factor (\textit{multi}), which defaults to 3 for our experiments.
            DTW$_{\text{Single}}$ alignments are then calculated between each subset $A_i$ and target $T$, leading to a variation from \cref{eq:dtw_single} for $multi > 1$: 
            \begin{equation}
                \begin{aligned}
                    & \text{DTW}_{\text{Multi}}(A, T) = \star_{i=1}^{\textit{multi}} \text{DTW}_{\text{Single}}(A_i, T), \\
                    & \qquad \text{where } \star \in \{\min, \text{avg}\}, \\
                    & \qquad A_i = A\left[\left\lfloor \frac{(i-1) \cdot |A|}{\textit{multi}} \right\rfloor : \left\lfloor \frac{i \cdot |A|}{\textit{multi}} \right\rfloor \right], \\
                    & \qquad A[x:y] \text{ represents a segment from index } x \text{ to } y
                \end{aligned}
                \label{eq:dtw_multi}
            \end{equation}
            
            This produces the same matrix structure as DTW$_{\text{Single}}$ but aggregated over all subsets $A_i$.
            By dividing into smaller time frames, we reduces the signal deviations caused by variations inside longer time series possibly leading to a shorter but better alignment.
        
        \textbf{Slicing-DTW-Attack.}\label{sec:Slicing_DTW_Attack}
            In the \textit{Slicing-DTW-Attack}, we invert the idea of the Multi-DTW-Attack and instead divide the signals of target $T$ into slices matching the length of the attack sample $A$.
            For this approach we assume $|A| \leq |T|$ .
            For better coverage, we create an overlap of 50\% between slices using a sliding window.
            We can then perfectly align the resulting target slices with the same length attack sample for calculating the DTW$_{\text{Single}}$ distance between the signals, which removes length-related differences and better focuses on the similarity of specific moments in the time series.
            Distances are calculated incorporating \cref{eq:dtw_single}:
            \begin{equation}
                \begin{aligned}
                    & \text{DTW}_{\text{Slicing}}(A, T) = \star_{j=1}^{\textit{slices}} \text{DTW}_{\text{Single}}(A, T_j), \\
                    & \qquad \text{where } \star \in \{\min, \text{avg}\}, \quad \textit{slices} = \left\lceil \frac{2 \cdot |T|}{|A|} \right\rceil, \\
                    & \qquad T_j = T\left[\left\lfloor \frac{(j-1) \cdot |A|}{2} \right\rfloor : \left\lfloor \frac{(j-1) \cdot |A|}{2} + |A| \right\rfloor \right], \\
                    & \qquad T[x:y] \text{ represents a segment from index } x \text{ to } y
                \end{aligned}
                \label{eq:dtw_slicing}
            \end{equation}
            
            This again generates the same matrix structure as DTW$_{\text{Single}}$ but aggregated over all slices $T_j$.
        
        \textbf{Multi-Slicing-DTW-Attack.}\label{sec:Multi_Slicing_DTW_Attack}
            Finally, the \textit{Multi-Slicing-DTW-Attack} combines the Multi-DTW-Attack and Slicing-DTW-Attack.
            Thus, the attack sample $A$ is divided into subsets as in \cref{eq:dtw_multi} and the dataset signals are sliced using \cref{eq:dtw_slicing}, resulting in:
            \begin{equation}
                \begin{aligned}
                    & \text{DTW}_{\text{Multi-Slicing}}(A, T) = \star_{i=1}^{\textit{multi}} \text{DTW}_{\text{Slicing}}(A_i, T), \\
                    & \qquad \text{where } \star \in \{\min, \text{avg}\}, \\
                    & \qquad A_i = A\left[\left\lfloor \frac{(i-1) \cdot |A|}{\textit{multi}} \right\rfloor : \left\lfloor \frac{i \cdot |A|}{\textit{multi}} \right\rfloor \right], \\
                    & \qquad A[x:y] \text{ represents a segment from index } x \text{ to } y
                \end{aligned}
                \label{eq:dtw_multi_slicing}
            \end{equation}
            
            We aggregate slices $T_j$ and subsets $A_i$, keeping the $n \times m$ matrix.
            
        \textbf{Aggregation.}
            For step (6) in our attack framework from \cref{fig:framework-overview}, we focus on the aggregation function $\star$ from \cref{eq:dtw_multi,eq:dtw_slicing,eq:dtw_multi_slicing}.
            In these equations, we state that $\star \in \{\min, \text{avg}\}$, which is needed due to the multiple scores per target returned by our multi and slicing strategies.
            For our ranking, these scores are combined into a single value per sensor modality $m \in M$ using one of two methods: $\text{avg}$, which computes the mean distance score for each modality, or $\min$, which selects the minimum distance score for each modality.
            For Multi-Slicing-DTW-Attacks, we need to decide on a combination between these methods.
            Our tuning experiments show that the $\text{avg}$ best suits the Multi-DTW-Attack, the $\min$ excels for Slicing-DTW-Attack, and $\min$-$\min$ for Multi-Slicing-DTW-Attack, as can be seen in our results in \cref{app:results} \cref{app:tab:comparision_min_avg}.
            We therefore use these methods in further experiments.
            We address the aggregation of multiple sensor modalities in \cref{sec:Sensor Ranking}.
            
    \subsection{Rank-based Evaluation}\label{sec:ranking}
    
        We want to define a rank-based metric that allows us to assess attack success beyond direct re-identification, such as retrieving the correct target $T_A$ within the top five distance scores.
        Correctly putting a target into the top ranking spots might still pose a privacy risk, due to reducing the search space for further exploits.
        
        We convert the DTW distances between an attack sample $A$ and targets $\mathcal{D}_T$, where each target $T \in \mathcal{D}_T$ gets associated with a single distance score $d$.
        These $(T:d)$ pairs are sorted in ascending order based on $d$ in the list $R$, assigning ranks based on the smallest distance.
        We employ realistic rank selection for ties based on \citet{berrendorf2020ambiguity}.
        With $R$ and $\alpha = d_{T_A},$ where $(T_A:d_{T_A}) \in R$ is the distance score of the correct target $T_A$, the realistic rank is the arithmetic mean of the optimistic rank ($rank^+(R, \alpha) = |\{\beta \in R \mid \beta > \alpha\}| + 1$) and the pessimistic rank ($rank^-(R, \alpha) = |\{\beta \in R \mid \beta \geq \alpha\}|$), resulting in: $Rank(R, \alpha) = \frac{1}{2} (rank^+(R, \alpha) + rank^-(R, \alpha)) \label{eq:realistic-rank}.$
            
        Based on these ranks, we calculate the precision@k (p@k) scores regarding the correct target $T_A$ and $|\max(k)=\mathcal{D}_T|$, which quantifies $T_A$ among top-k ranks.
        Thus a p@1 gives our success for direct re-identification by ranking the correct target at the top.
        We also consider a higher k-value in p@5, since even if we are not able to rank the target first, we might still be able to significantly narrow down the list of candidates to the top five.

    \subsection{Evaluation Pipeline}\label{sec:Evaluation Pipeline}
        For tuning our DTW attack to the optimal parameters in each run, we go through a pipeline comprising three evaluation phases: class ranking, sensor aggregation, and attack window sizes (see (7) of \cref{fig:framework-overview}).
        But before starting our evaluations for attack hyperparamters, we need to consider the simulation strategy regarding target selection.
        We aim to adhere to an attack scenario in which we individually target a subject for whom we have obtained a sample $A$ and their corresponding target data $T_A$ within the dataset $\mathcal{D}_T$.
        However, we average our results over multiple subjects in our target datasets for a subject-based cross-validation approach.
        Thus, e.g., using the WESAD dataset of 15 subjects as $\mathcal{D}_T$, each individual subject is a target $T$ and attacked once in a simulation from \cref{alg:simulation}.
        Thus, $|\mathcal{D}_A| = |\mathcal{D}_T|$.
        The returned p@k value lists $P$ for each simulation are averaged over the resulting set of attack samples $\mathcal{D}_A$ for an overall evaluation: $\overline{P}_{\mathcal{D}_T} = \frac{\sum_{A \in \mathcal{D}_A} P_A}{|\{A \in \mathcal{D}_A\}|}$.
        
        Our synthetic datasets with the same size (CGAN$_{15}$ and DGAN$_{15}$) also get the same handling.
        The bigger datasets are evaluated by adding subjects to $\mathcal{D}_T$ but keeping the same 15 targets for $\mathcal{D}_A$, which can lead to $|\mathcal{D}_A| < |\mathcal{D}_T|$.
        So e.g., for evaluating a CGAN$_{100}$, we would take the same 15 targets from the CGAN$_{15}$ but add another 85 synthetic subjects to increase the retrieval difficulty.
        
        For our standard case we assume the worst case $\forall A \in \mathcal{D}_A: T_A \in \mathcal{D}_T$, i.e., the correct target $T_A$ for an attack sample $A$ is always included in the target dataset $\mathcal{D}_T$.
        In \cref{sec:in-out}, we loosen that assumption to also adhere to the partial $\mathcal{M}$ from \cref{def:reidentification}.

        \subsubsection{Classes}\label{sec:Classes}
            First, we consider the influence of classes on our attack. 
            We start by evaluating the potential for stress detection to enhance our precision, since models reliably classify both classes in our data~\cite{GilMartin2022HumanStress,Lange2024GANStressDetection}.
            With the assumption that we can classify the attack sample and the target regarding the two classes, we can try and optimize by comparing only the same types.
            In tuning experiments on the WESAD dataset, we evaluated this scenario and found the non-stress data to have an average advantage of almost 29\% over stress data for re-identification in all attacks except the Slicing-Attack, where we found no difference since both classes hit the maximum.
            See \cref{app:tab:classes} of \cref{app:results} for the detailed results.
            
            For our general attack results, however, we want to give a more realistic view without such classification pre-requisites, for which we have to equally consider both classes in our datasets.
            With a class distribution of 70\% non-stress and 30\% stress inside the data, we generally perform two separate attacks using corresponding attack samples cut from each class as described in the simulation mode from \cref{sec:data model}.
            We then combine their p@k scores using a weighted mean to reflect the class prevalence: $\mathcal{W}_{classes}(n,r_{non},r_{stress}) = \quad (0.7 \cdot \text{p@k}(r_{non}) + 0.3 \cdot \text{p@k}(r_{stress}) | k \in (1,2,...,n) )$, where $n=|\mathcal{D}_T|$ and $r_{non},r_{stress}$ the ranks of the correct target $T_A$ returned by the attacks for each class.
            With this approach we remove potential over- or undervaluation of the attack results based on the class ratio or type of class inside the the attack sample.

        \subsubsection{Sensor Aggregation}
            For the subsequent second stage, we identify the best aggregation method for the multiple sensor-level distance scores into our ranking from \cref{sec:ranking}.
            
            \textit{Naive Sensor Aggregation.}\label{sec:Sensor Ranking}
                This basic approach assumes a "naive" attacker who aggregates all sensor distances without selecting specific sensors.
                We calculate the mean across all sensor modalities $M=$\{BVP, EDA, TEMP, ACC\} in the DTW distance matrix $D$ as: $Agg(D) = \frac{\sum_{m \in M} D_m}{|M|}$, to obtain a general overview of attack performance, independent of individual sensors.
                However, this simple approach does not account for potential differences in re-identification effectiveness across sensor combinations.

            \textit{Informed Sensor Aggregation.}\label{sec:informed_sensor}
                With knowledge about the underlying data, DTW attacks can be enhanced by adapting the ranking to the importance of available sensor modalities.
                As an estimation, an attacker might create a priori sensor rankings based on the anonymized dataset, evaluating the most effective sensors by assuming random subject identifiers for each sample and splitting data as in the simulation mode (\cref{sec:data model}).
                Another option is to use synthetic data, allowing the attacker to identify and apply an informed sensor combination to the real dataset.

                Thus, to move beyond the naive approach, we evaluate each sensor and their combinations, averaging their combined distance scores as before.
                We evaluate all permutations of $M$ regarding $Agg(D)$ and select the one with the best p@1 score; if this is not possible, a random choice is made.
                In \cref{app:tab:sensor_rank} of \cref{app:results}, we present these permutation results for the WESAD dataset.
                The Single-, Multi-, and Multi-Slicing-Attacks strongly favor BVP, while the Slicing-Attack shows just minimal performance differences between combinations.
                We can see how averaging all sensors also gives a rough average of sensor combinations but just sub-optimal re-identification, except for the Slicing-Attack.
                Overall, BVP and ACC seem to pose the highest re-identification risks.
                
            \begin{figure}[ht]
                \centering
                  \includegraphics[width=0.42\linewidth]{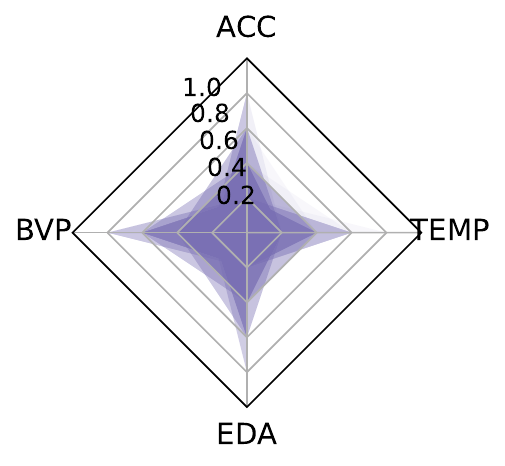}
                  \includegraphics[width=0.42\linewidth]{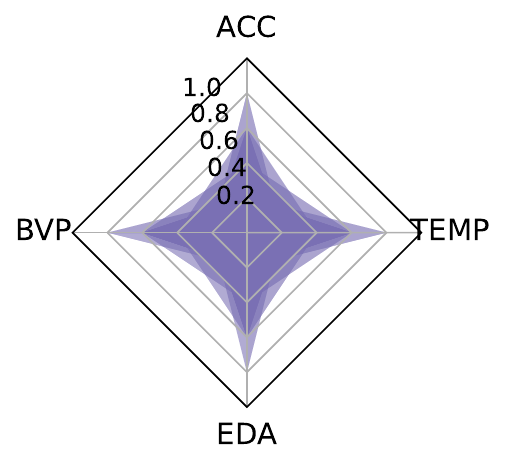}
                \caption{Radar charts illustrating the best sensor weightings for the Single-DTW-Attack, Multi-DTW-Attack and Multi-Slicing-DTW-Attack (left) and the Slicing-DTW-Attack (right). We layered all weightings delivering the best results.}
                \label{fig:sensor_weights}
                \Description{The differences between attacks and settings are conveyed in the main text.}
            \end{figure}
            
            \textit{Weighted Sensor Aggregation.}\label{sec:Weighted Sensor Ranking}
                We further determine optimal sensor weightings through grid search.
                Instead of equally weighted transmutations, we find specific weights $w$ for each sensor to derive a weighted mean: $Agg(D,W) = \frac{\sum_{m \in M} w_m D_m}{\sum_{m \in M} w_m}$, for a set of weights $W=\{w_m | m \in M\}$.
                Using equal or single weights leads to the naive and informed methods, respectively.
                In \cref{fig:sensor_weights}, the highest weightings still favor BVP and ACC, but now also EDA, in all but the Slicing-Attack, which shows no clear preference.
                We find:
                \begin{equation}
                    W_{opt} = \{ w_{\text{BVP}}: 0.4, \, w_{\text{EDA}}: 0.2, \, w_{\text{TEMP}}: 0.2, \, w_{\text{ACC}}: 0.2 \},
                	\label{eq:sensor_weighting}
                \end{equation}
                    as the only optimal weighting that applies across WESAD, CGAN$_{15}$, DGAN$_{15}$, all DTW-Attacks, and both classes.
                However, this scenario represents a worst-case for our multimodal attack and requires prior knowledge on the target data, so we consider it separately.

        \subsubsection{Attack Window Sizes}\label{sec:Attack Window Sizes}
            We test varying window sizes $w_A$ for our attack sample $A$ to assess the impact on re-identification performance.
            For Single-DTW- and Slicing-DTW-Attacks, we evaluate window sizes from 1,000 (1) to 36,000 (36) data points, while the Multi-DTW-Attacks take the same maximum but split them into three separate windows of up to 12,000 (12) each.
            One window translates to $\approx$16 seconds.
            In a first test on the WESAD dataset (see \cref{app:fig:evaluation_attack_window_sizes} in \cref{app:results}), we find that lower sizes are generally better in all attacks except the Slicing-Attack, where we find the same results for all sizes.
            We even observe that additional window size does hinder performance, especially for the Multi-Slicing-Attack.
            The advantage of less windows in our other attacks is an intuition that also motivates the slicing approach by suggesting that comparing shorter time series further enhances individual assignment and re-identification performance on our data.
            For our attack results, we test all window sizes and then select the smallest best performing size for each attack to measure the worst-case.
    
    \subsection{In-Out Threshold Scenario}\label{sec:in-out}
        As described in \cref{sec:Evaluation Pipeline}, we have until now retrieved correct targets for attack samples under the optimistic assumption that they are always included in the target dataset: $\forall A \in \mathcal{D}_A: T_A \in \mathcal{D}_T$.
        This approach allowed us to focus on varying the retrieval difficulty by adjusting the size of $\mathcal{D}_T$ but represents an upper limit of our re-identification risk.
        However, to create a more realistic picture of attack performance, we also consider an "in-out" attack scenario.
        In this scenario, only a fraction of available attack samples may have corresponding targets within the dataset: $\forall A \in \mathcal{D}_A: T_A \in \mathcal{D}_T \vee T_A \notin \mathcal{D}_T$.
        Consequently, the attacker does not know whether the correct target for an attack sample is present or not.
        
        To achieve a simulation of this partition, we use two separate datasets: $\mathbb{D}_A$ (attack subjects) and $\mathbb{D}_T$ (target subjects), each with the same number of subjects ($|\mathbb{D}_A| = |\mathbb{D}_T|$).
        We define a percentage overlap as the Dice coefficient $DSC = \frac{ 2 \cdot |\mathbb{D}_A \cap \mathbb{D}_T|}{|\mathbb{D}_A| + |\mathbb{D}_T|}$ between the datasets, representing the proportion of possible correct matches.
        The remaining subjects are non-overlapping and should not be matched.
        For the WESAD dataset, which contains 15 subjects, the lowest possible overlap is 1 subject, translating to $DSC_{L} = 12.5\%$, defining our low overlap setting.
        We also employ a medium overlap of $DSC_{M} = 50\%$ and a full overlap, or $DSC_{F} = 100\%$, which represents our standard scenario.
        For synthetic GAN datasets, we apply these overlaps to larger dataset sizes of $|\mathbb{D}_A| = |\mathbb{D}_A| = 120$, filling them according to each overlap requirement.
        
        For adapting our attacks to the "in-out" problem, we introduce an additional threshold stage based on the DTW distances.
        This stage determines whether a returned subject is sufficiently close in distance to the attack sample to potentially be the correct target.
        For incorporating this threshold $\mathscr{t}$ into our attack, we have to revise \cref{alg:attack} to its threshold variation, which now also fits the partial setting of $\mathcal{M}$ from \cref{def:reidentification}.
        
        We can evaluate our threshold attack utilizing the same simulation techniques (\cref{sec:data model}) as before by treating $\mathbb{D}_A$ and $\mathbb{D}_T$ as $\mathcal{D}_A$ and $\mathcal{D}_T$.
        We track the correct and incorrect matches between $\mathbb{D}_A$ and $\mathbb{D}_T$ using p@1, recall@1 (r@1), and their combined F1@1-score (F@1) across various thresholds ($\mathscr{t}$).
        Our F1 translates to F@1, since our evaluation is not limited to deciding between "in" and "out" but instead also considers if the found match is correct or not.
    
    \subsection{Mitigating DTW-Attacks}\label{sec:privacy}
        \begin{figure}[ht]
          \centering
          \includegraphics[width=1\linewidth]{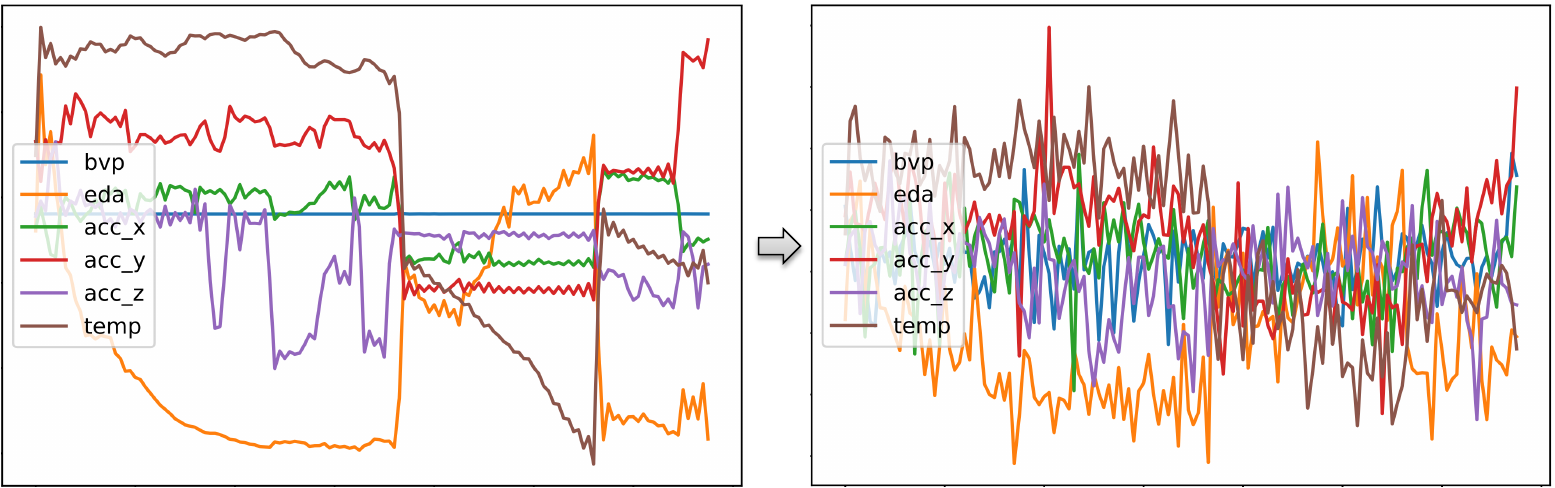}
          \caption{Example effects of adding random noise using the Laplace distribution with $\sigma=0.1$ to subject s$_2$ of the WESAD dataset, which was downsampled using a $DSF=1000$.}
          \label{fig:noisy_data}
          \Description{Left: normal signal visualization for the modalities. Right: the signals are clearly distorted by the noise, exhibiting amplitude changes at every step.}
        \end{figure}
        
        To thoroughly assess our attacks, we also test the effectiveness of defense mechanisms against them.
        As noted in \cref{sec:intro}, data owners commonly use de-identification for privacy protection.
        Our attacks, however, target de-identified health data without metadata, underscoring the need for additional data-centric defenses.
        
        One common approach is noise injection, as seen in Differential Privacy (DP)~\cite{dwork2006differential}, which is able to obscure biometric features while preserving analytical utility.
        This method adds random noise to the original data, which is drawn from a Laplace distribution~\cite{Kotz2001Laplace}.
        For our time series data, we inject noise by varying the scale parameter $\sigma$, or noise multiplier, which controls the magnitude.
        A larger $\sigma$ introduces more noise to data points, potentially reducing susceptibility to attacks but also impacting data utility.
        \Cref{fig:noisy_data} illustrates how noise affects WESAD data, revealing significant signal distortion even at low noise levels ($\sigma=0.1$).
        
        To evaluate the privacy-utility trade-off, we conduct a binary stress detection, employing a convolutional neural network model and methodology from previous work~\cite{GilMartin2022HumanStress, Lange2024GANStressDetection}.
        We apply noise levels ranging from $\sigma=[0,15]$ to the WESAD dataset creating noisy versions, where $\sigma=0$ represents no noise.
        We then run our attacks and the stress detection model on the noised data, however not adding noise to the attack samples.
        Attack performance is measured using p@1, while the utility of the stress detection is given by F1-score via leave-one-out cross-validation.
        Both metrics are averaged over 10 repetitions to account for the introduced randomness.
        Our goal is to find a noise level that provides protection against our attacks, while still allowing for meaningful data analysis.
        
\section{Evaluation}\label{sec:evaluation}
    In this section, we present the results of our propsoed experiments.
    \begin{figure*}[ht]
        \centering
            \includegraphics[trim={1in 0.27in 1in 1in},clip,width=1\linewidth]{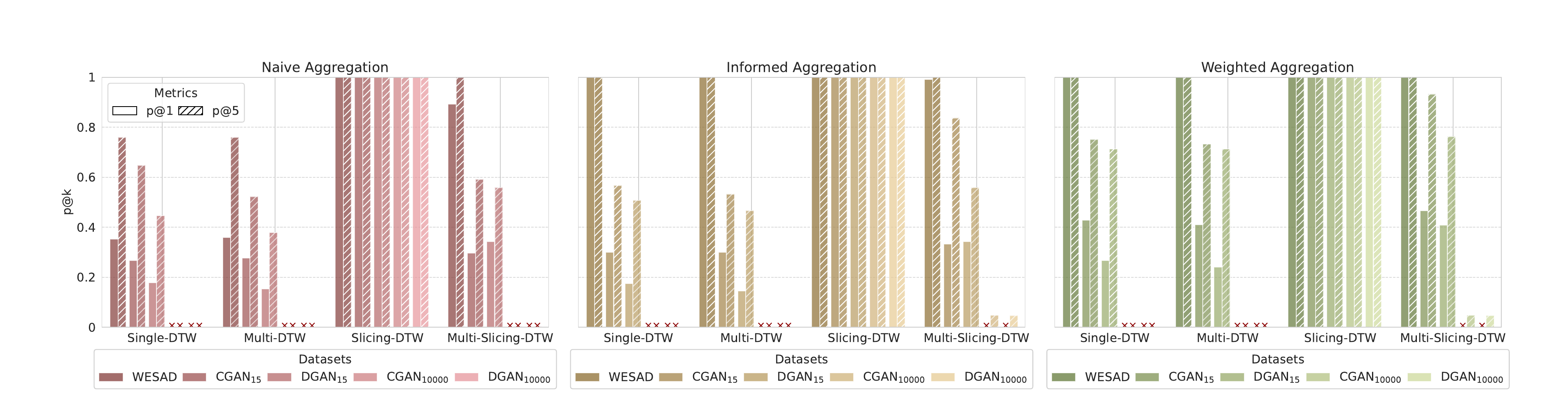}
        \caption{Our attack results (p@1, p@5) in the standard evaluation setting. We test our four DTW attacks on the WESAD dataset and on synthetic datasets of sizes 15 and 10,000, all using a $DSF=1000$. Results that are effectively giving zero are indicated with red crosses, which in these cases translates to random guessing. Additional details, including the attack window size $w$ and the best-performing sensors for informed aggregation, are provided in \cref{app:tab:attacks_eval} of \cref{app:results}.}
        \label{fig:attacks_eval}
        \Description{The differences between attacks and settings are conveyed in the main text.}
    \end{figure*}
    
    \subsection{Standard DTW-Attack Results}\label{sec:evaluation_standard}
    In this first evaluation, we focus on main attack scenario (\cref{sec:Evaluation Pipeline}) using the different sensor aggregation from \cref{sec:Sensor Ranking}.
    We focus on the mean over all signals (naive) and picking sensor combinations (informed) as the realistic cases, since finding optimized weights (weighted) requires more detailed a priori knowledge by the attacker.
    The naive approach allows for comparing the general re-identification potential of our attacks, while the informed variation reflects a more pessimistic threat level, and the weighted version represents the worst-case scenario.
            
        \textbf{Complexity Reduction.}
            To evaluate the complexity reduction methods from \cref{sec:Complexity Reduction}, we compare the average results across all naive attacks and the WESAD, CGAN$_{15}$, and DGAN$_{15}$ datasets.
            Our focus is on the usability of the reduced data for our attacks.
            Detailed results are located in \cref{app:tab:complexity} of \cref{app:results}.

            For downsampling, we test $DSF$s ranging from $1\times$ (no downsampling) to $1000\times$ reduction.
            We observe a linear relationship, where a higher $DSF$ leads to lower runtimes and improved results, even compared to the original data.
            This improvement is likely due to signal smoothing, which may aid DTW alignment.
            From $DSF=1$ to $DSF=1000$, we see an average 21\% increase in p@1 across datasets.
            DBA and PCA are evaluated on datasets downsampled with the optimal $DSF=1000$.
            While both methods reduce execution time, they significantly decrease re-identification success.
            Compared to downsampling alone, DBA results in an average p@1 loss of 30\%, and PCA results in a 65\% loss.
        
        \textbf{Naive Aggregation.}
            Our attack results are summarized in \cref{fig:attacks_eval}, where we first focus on the naive aggregation method.
            The Slicing-DTW-Attack consistently performs best across all datasets, achieving perfect scores for both WESAD and the synthetic dataset with 15 subjects, and even for the larger 10,000-subject datasets.
            No other attack produced results better than random guessing ($1/10000 \approx 0.01\%$) on these large datasets.
            When comparing the attacks on 15-subject datasets, Single-DTW performed the worst, followed by the Multi-Attack.
            The Multi-Slicing variant performed better than both and came close to the Slicing variant, achieving 89\% on WESAD.
            Outside of the Slicing-Attack, we observe that CGAN poses a more challenging task than WESAD, with DGAN being even harder for most attacks, except for the Multi-Slicing, where the trend is reversed.
            Despite this, all attacks exceeded the random guessing baseline of $1/15 \approx 6.7\%$ for p@1.
            When considering p@5 results, our attacks are significantly stronger, effectively reducing the candidate pool to the top five.
            In \cref{app:tab:attacks_eval} of \cref{app:results}, we provide additional details, including the attack window sizes $w$.
            Notably, Slicing-DTW uses the smallest window size of $w=1$ for the 15-subject datasets (16 seconds), while nearly maximizing the size for the larger datasets at $w=34$ and $w=32$ (about 9 minutes).

        \textbf{Informed Aggregation.}
            Our informed sensor aggregation offers a more refined approach to handle multimodal data in our attacks, as certain sensors may be more effective for specific attacks.
            Selecting sensor combinations proves highly effective on the WESAD dataset, with all attacks achieving maximum p@1, except for Multi-Slicing, which closely follows at 99\%.
            While Single- and Multi-Attacks don’t show significant improvements on the GAN datasets (1\%).
            Notably, Multi-Slicing-DTW surpasses random guessing on these datasets and significantly improves its p@5 score on CGAN$_{15}$.
            For the Slicing-Attack, we do not observe a distinct advantage with the informed aggregation, as it already achieved perfect results using the naive approach.
            The selected sensor combinations are detailed in \cref{app:tab:attacks_eval} of \cref{app:results}.
            While Single-, Multi-Attack, and Multi-Slicing favor BVP, the Slicing-Attack performs well across most combinations, particularly excluding the sole use of BVP.

        \textbf{Weighted Aggregation.}
            Compared to the naive and informed aggregations, the weighted aggregation represents a worst-case scenario due to the significant a priori knowledge required to determine optimal sensor weights.
            Using the optimal weighting $W_{opt}$ from \cref{eq:sensor_weighting}, we observe substantial improvements across all attacks, particularly on the smaller synthetic datasets.
            On the 15-subject GAN datasets, we find an average p@1 increase of 12\% for the Single-, Multi-, and Multi-Slicing attacks compared to the naive approach.
            While WESAD had already achieved maximum p@1 for all attacks except Multi-Slicing, this method now reaches the maximum as well.
            However, we cannot surpass our informed results on the 10,000-subject datasets.
            The Slicing-Attack again sees no offers no noticeable improvement with weighted aggregation due to its already maxed out results.
            
    \begin{figure*}[ht]
        \centering
        \includegraphics[width=0.24\linewidth]{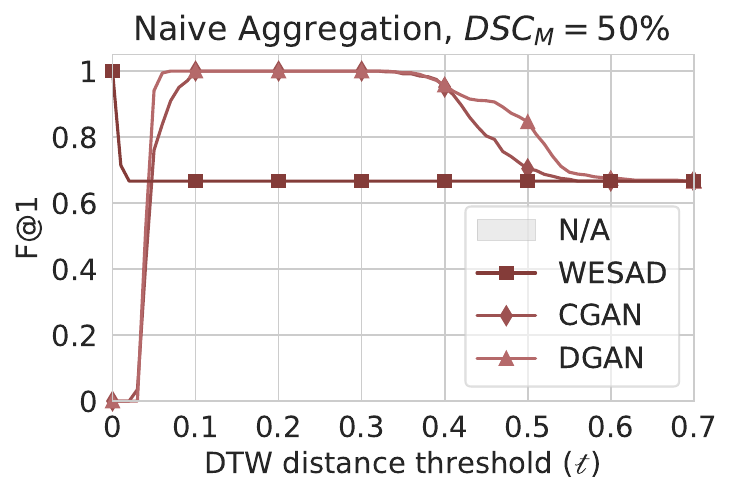} 
        \includegraphics[width=0.24\linewidth]{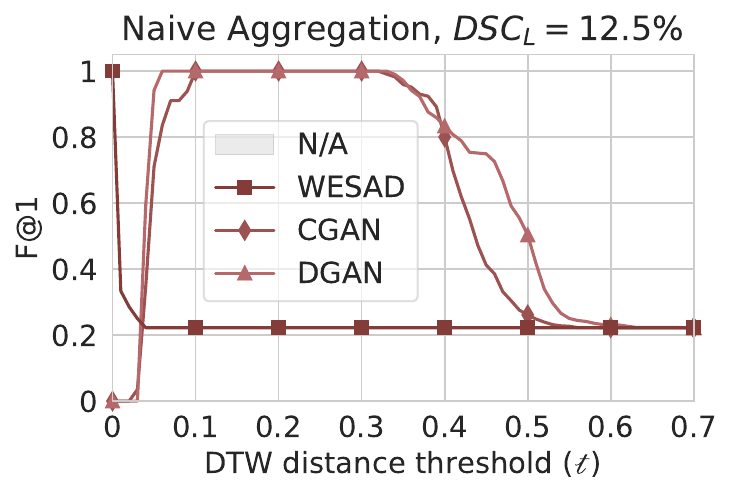}
        \includegraphics[width=0.24\linewidth]{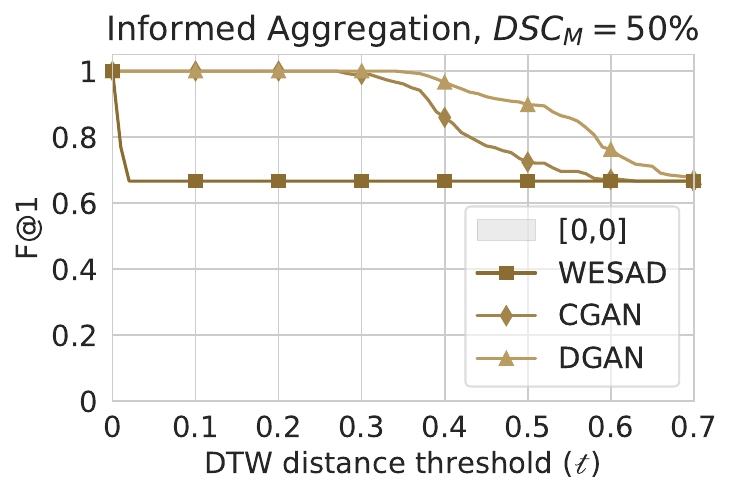}
        \includegraphics[width=0.24\linewidth]{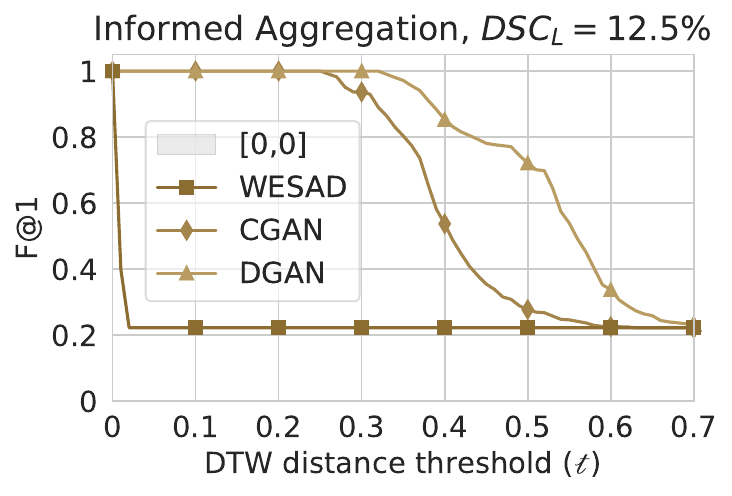}
        \caption{Evaluation of the "in-out" scenario on varying overlaps. The x-axis gives the applied distance threshold and the y-axis shows the F@1-score for our Slicing-DTW-Attack. We highlight the optimal threshold range across the tested datasets.}
        \label{fig:evaluation_in_out}
        \Description{The differences between attacks and settings are conveyed in the main text.}
    \end{figure*}
    
    \begin{figure}[ht]
        \centering
        \includegraphics[width=0.49\linewidth]{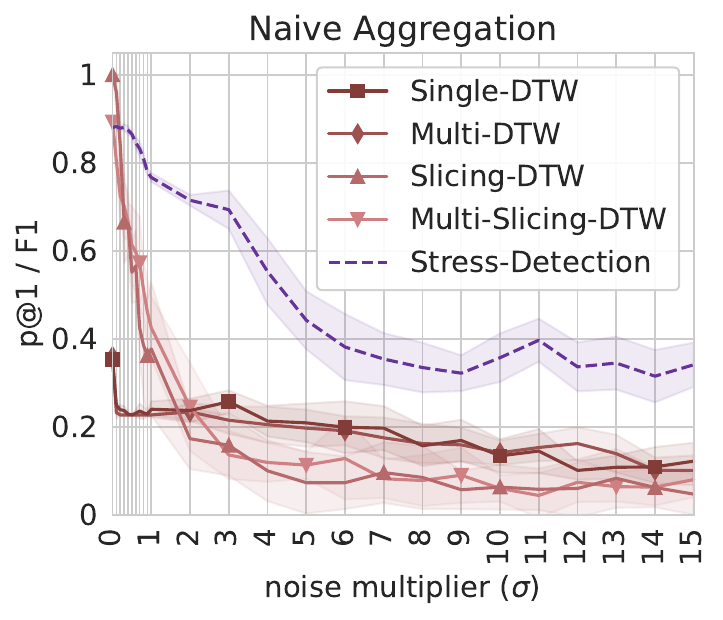}
        \includegraphics[width=0.49\linewidth]{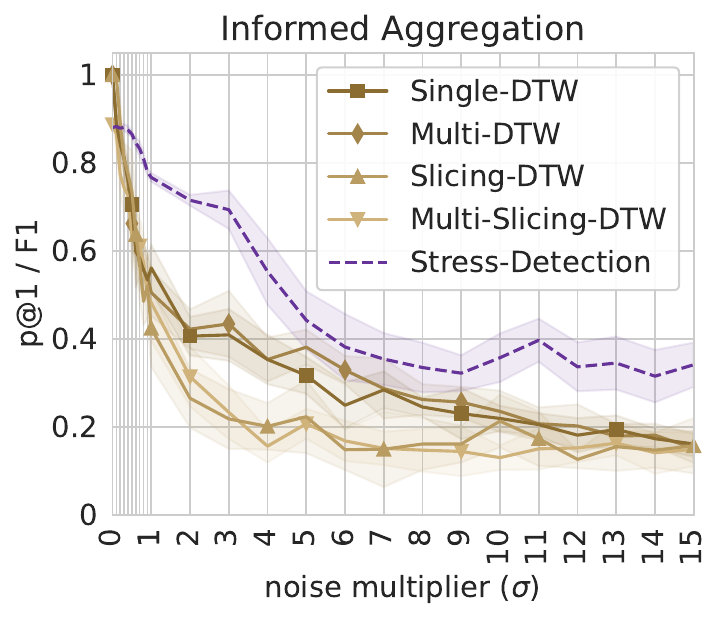}
        \caption{Evaluation of the privacy-utility trade-off from noise injection into the WESAD data. The y-axis shows the average attack p@1 and stress detection F1-score over 10 runs, with the x-axis giving the applied noise multiplier in $\sigma=[0,15]$ with steps of 0.1 between [0,1] and 1 thereafter.}
        \label{fig:evaluation_privacy_usability}
        \Description{The differences between attacks and settings are conveyed in the main text.}
    \end{figure}

        \textbf{Runtime.}
            For discussing runtime scalability, we focus on one attack run using one attack sample under the external mode from \cref{sec:data model} and use our parameters from the best naively aggregated attack results.
            The scalability is particularly supported by the $DSF=1000$.
            We use the Python package DTAIDistance with the \textit{distance\_fast} method by \citet{Meert2020dtaidistance}. 
            Since we do not rely on early stopping optimizations from e.g., \citet{Rakthanmanon2021dtw}, we only achieve a DTW time complexity of $O(n^2)$ regarding the length of the compared time series.
            Since DTW runtime scales with the time series lengths, it is independent of our general attack framework and instead fully dependent on the underlying data.
            
            Assuming similar time series lengths, as in our case, the attacks itself scale linearly ($O(n)$) with an increasing subject count in the target dataset.
            The DTW calculations are the primary contributors to runtime, while the duration for ranking heavily depends on dataset size, as detailed in \cref{app:tab:evaluation_runtime} of \cref{app:results}.
            Our most effective attack, the Slicing-DTW-Attack, can target a dataset of 10,000 subjects, each with approximately 138,000 data points, in about 2 minutes—a reasonable timeframe.
            The fastest attack, the Multi-DTW-Attack, completes the same task in under 70 seconds.
            
            For these results, we used a single core of an AMD EPYC 7551P CPU, indicating potential for further speedup through parallelization across multiple cores or machines.
            For multiple attack samples, runtime scales quadratically ($O(n^2)$), as each sample requires an attack on the dataset.
            Parallel computation should be preferred to maintain feasible runtimes in such cases.
    
    \subsection{In-Out Threshold Results}\label{sec:eval_in_out}
    
        The results of our "in-out" experiments, detailed in \cref{sec:in-out}, are shown in \cref{fig:evaluation_in_out}.
        We use the Slicing-DTW-Attack, which performed best in our standard evaluation (\cref{sec:evaluation_standard}).
        For sensor aggregation, we compare only the naive and informed methods, as the weighted strategy requires excessive prior knowledge.
        Each dataset’s F@1 score is presented across thresholds $\mathscr{t} = [0, 0.7]$ for two overlap settings: medium ($DSC_M$) and low ($DSC_L$).
        While $\mathscr{t} = 0.7$ is not the maximum threshold (since DTW distance is uncapped), the most critical results for our datasets occur at these thresholds.
        Full overlap ($DSC_F$) results are provided in \cref{app:fig:evaluation_in_out} in \cref{app:results}, as they align with the standard evaluation in \cref{sec:evaluation_standard} for $\mathscr{t} \geq 0.1$ and $\mathscr{t} \geq 0$ for the naive and informed settings, respectively.
        
        Notably, the WESAD dataset consistently achieves maximum F@1 at the minimum DTW distance threshold $\mathscr{t} = 0$.
        This behavior is mirrored by the GAN datasets in the informed setting, although they peak later than WESAD in the naive setting.
        A consistent feature across all settings is that WESAD’s F@1 drops sharply for $\mathscr{t} > 0$, quickly stabilizing at a plateau relative to the overlap.
        The GAN datasets, however, exhibit a broader optimal threshold range, followed by a steady, less pronounced decline.
        Among these, the DGAN dataset shows the best results across thresholds, indicating that WESAD distances are significantly closer and better aligned than the synthetic GAN datasets, which may explain the higher difficulty in retrieving them in our attack evaluation (\cref{sec:evaluation_standard}).
        In terms of common threshold ranges, only the GAN datasets align in the naive aggregation at $\mathscr{t} = [0.1, 0.32]$.
        In the informed case, all three datasets reach optimal performance at $\mathscr{t} = 0$, after which WESAD drops off, leaving the GAN datasets to maintain their range until $\mathscr{t} = 0.27$ for $DSC_M$ and $\mathscr{t} = 0.25$ for $DSM_L$.
        Overall, we identify thresholds that deliver optimal F@1 scores for each dataset across all overlap and aggregation settings, highlighting the robustness of our attack and ranking methodology, even in scenarios with substantial non matches.
        However, a common threshold across all datasets is only found with the informed aggregation strategy.
        
        A limitation of the threshold-based approach is determining the correct threshold without direct access to the dataset. However, leveraging insights from the record linkage domain~\cite{franke2024quality}, we can use a privacy-conscious a priori estimation of linkage quality to guide threshold selection.
    
    \subsection{Mitigation Results}\label{sec:evaluation_privacy}
    
        In our defense experiment, introduced in \cref{sec:privacy}, we demonstrate the effectiveness of noise injection in reducing re-identification risk from DTW attacks while preserving data utility. This is exemplified by a stress detection task on the WESAD dataset, which showed especially prone to our attacks in \cref{sec:evaluation_standard}. The results are given in \cref{fig:evaluation_privacy_usability}, where utility is measured by the F1-score of a stress detection model and privacy risk by the p@1 of our DTW attacks, using naive and informed aggregations. Both metrics are evaluated across noise multipliers $\sigma = [0, 15]$.

        Stress detection proves resilient to low noise levels, with the F1 decreasing just slightly from 0.88 (no noise, $\sigma = 0$) to 0.87 at $\sigma = 0.5$. Utility then declines noticeably but stabilizes around 0.7 between $\sigma = [2, 3]$. Beyond this point, F1 continues to drop, eventually converging to 0.3, the equivalent to random guessing stress labels.
        
        In contrast, re-identification risk from DTW attacks drops sharply at low noise noise levels, across both aggregation methods. All attacks show a steep initial decline up to $\sigma = 1$, after which p@1 gradually converges to under 0.16 at $\sigma = 15$. The Slicing-DTW-Attack, the most effective from our evaluation (\cref{sec:evaluation_standard}), drops from perfect re-identification (p@1 = 1) to 0.21 (naive) and 0.42 (informed) at $\sigma = 1$, continuing to decline more rapidly than the Single- and Multi-DTW attacks. These latter attacks might be more robust to noise, likely due to its disruption of smaller alignments, diminishing the advantage of Slicing attacks. Further, both attacks show to be significantly more threatening in the informed setting.
        
        The key takeaway is that noise can significantly reduce re-identification risk with relatively lower impact on usability. For example, at $\sigma = 2$, stress detection retains a relatively high F1-score of 0.71 (a 17\% decrease), while the best attack’s p@1 drops to 0.25 in the naive (a 74\% decrease) and 0.42 in the informed setting (a 58\% decrease). However, we do not achieve complete prevention of re-identification; at $\sigma = 15$, p@1 only drops to 0.12 for naive and 0.16 for informed, though stress detection becomes unusable. Still, our defense is particularly effective against the Slicing approach and generally suggests that noise injection in time series data can to some extent balance privacy and utility in mitigating DTW attacks.
 
\section{Discussion}\label{sec:discussion}
We introduced a modular framework for DTW re-identification attacks using novel approaches to enhance our success rates.
In this section, we discuss research questions (RQ) including limitations regarding our approach and findings from \cref{sec:evaluation}.

\noindent\textbf{RQ1: How severe is the actual threat from our DTW-Attacks?}
    \indent
    In \cref{sec:evaluation_standard}, we identify the Slicing-DTW-Attack as our most powerful approach, delivering perfect re-identification on WESAD, CGAN$_{15}$, DGAN$_{15}$, and even the larger 10,000-subject synthetic GAN datasets. Advanced sensor aggregation methods are unnecessary here, as the attack already succeeds under the naive aggregation strategy. While re-identification on the smaller datasets requires only 16-second long attack samples, this increases to 9-minute samples for the larger sets. Runtime stays at feasible levels across all tested datasets.
    This severe threat level is unmatched by other DTW attacks, although they still achieve high rates outside of the 10,000-subject cases. The p@k criterion further enables to potentially reduce candidate lists. Re-identification improves with informed aggregation and are highest in the weighted setting, though obtaining the optimal weights is more complex than selecting suitable sensors, making it a worst-case scenario.
    
    Recognizing that realistic attacks must account for cases where the target is not in the dataset, we tested DTW distance thresholds to distinguish true matches from non-matches in \cref{sec:eval_in_out}. We identified optimal thresholds across WESAD, CGAN, and DGAN data in the naive setting, with informed aggregation revealing a common threshold for all datasets.
    This enables perfect re-identification even in scenarios with many non-matches.

\noindent\textbf{RQ2: What are appropriate defenses against DTW-Attacks?}
    \indent
    Our attacks demonstrate that de-identification alone is insufficient to protect health data due to similarity-based attacks. For reducing their threat levels, we propose injecting random noise into such time series. As shown in \cref{sec:evaluation_privacy}, this approach can significantly reduce attack success by 58\%, while in comparison only mildly affecting usability for machine learning, with a 17\% decrease in performance. Despite this, we still consider de-identification a crucial component of user privacy protection.
    A theoretical privacy guarantee could be achieved by applying noise with DP~\cite{dwork2006differential}. Alternatively, k-anonymity~\cite{Sweeney2002K-Anonymity} could be considered, though its application to time series data is complex and often unsuitable~\cite{Shou2013TimeSeriesAnonymity}.

\noindent\textbf{RQ3: To what extent can we adapt to other data and tasks?}
    \indent
    While our study is to some extent limited to the smartwatch stress detection use case, the DTW attacks are generally transferable to other time series data, particularly when using naive aggregation. The informed and weighted approaches may require some prior data analysis but can still be adapted. Our results can also serve as a pre-trained attack model with optimized parameters for similar datasets, particularly those involving stress indicators~\cite{prajod2024stressor}.

    Further, our similarity ranking has practical applications beyond attacks. It can be used to link similar subjects, e.g. patients, enabling related treatments such as medication adjustment. By identifying similar individuals in a database, new subjects can benefit from tailored recommendations based on their comparable health profiles.

\noindent\textbf{RQ4: Limits of available and synthetic data for evaluation?}
    \indent
    The main limitation of our DTW attack evaluation is the limited availability of public health data, requiring us to rely on smaller domain-specific datasets like WESAD.
    To address scalability issues, we use synthetic GAN datasets, which help validate the performance of our attacks on larger sets.
    As demonstrated by \citet{Lange2024GANStressDetection}, these GANs preserve key characteristics of WESAD without directly replicating the original subjects.
    However, we observe greater variance in GAN sensor signals and different DTW distances between subjects compared to WESAD, as discussed in \cref{sec:datasets}.
    There is a limit to how much a GAN can diversify from such a small number of original subjects.
    Still, re-identification on the GAN datasets appears to be more challenging, potentially leading to an underestimation of the actual risk in real large-scale datasets.
    They thus provide a conservative estimate of re-identification risk.
         
\section{Conclusion}\label{sec:conclusion}
The collection of de-identified health time series data from smart devices, particularly smartwatches, is a common practice. However, such data contains biometric features that, despite the absence of direct identifiers, enable re-identification attacks. Our DTW-based attack methods demonstrate perfect re-identification by exploiting these inherent characteristics. Our Slicing-DTW-Attack proves especially effective, even on larger datasets. This confirms that de-identification alone is insufficient for identity protection. We further show that injecting random noise into the time series data can effectively reduce the success of our attacks while maintaining a reasonable utility-privacy trade-off for machine learning tasks.

Future work should focus on acquiring larger real-world datasets and applying this approach to other domains and tasks. 

\subsubsection*{Availability} 
Reference code for all experiments is available from our repository at \url{https://github.com/tobiasschreieder/dtw-attacks}. 

\subsubsection*{Ethical Principles}
Health data originated from public sources for research and was solely used within the limited scope of this work.

\begin{acks} 
We thank Nils Wenzlitschke and Florens Rohde for their methodical contributions.
The authors acknowledge the financial support by the Federal Ministry of Education and Research of Germany and by the Sächsische Staatsministerium für Wissenschaft, Kultur und Tourismus for ScaDS.AI.
Computations for this work were done (in part) using resources of the Leipzig University Computing Centre.
\end{acks}

\printbibliography
\appendix

\section{Additional Results}\label{app:results}
\begin{figure}[H]
        \centering
        \includegraphics[width=0.72\linewidth]{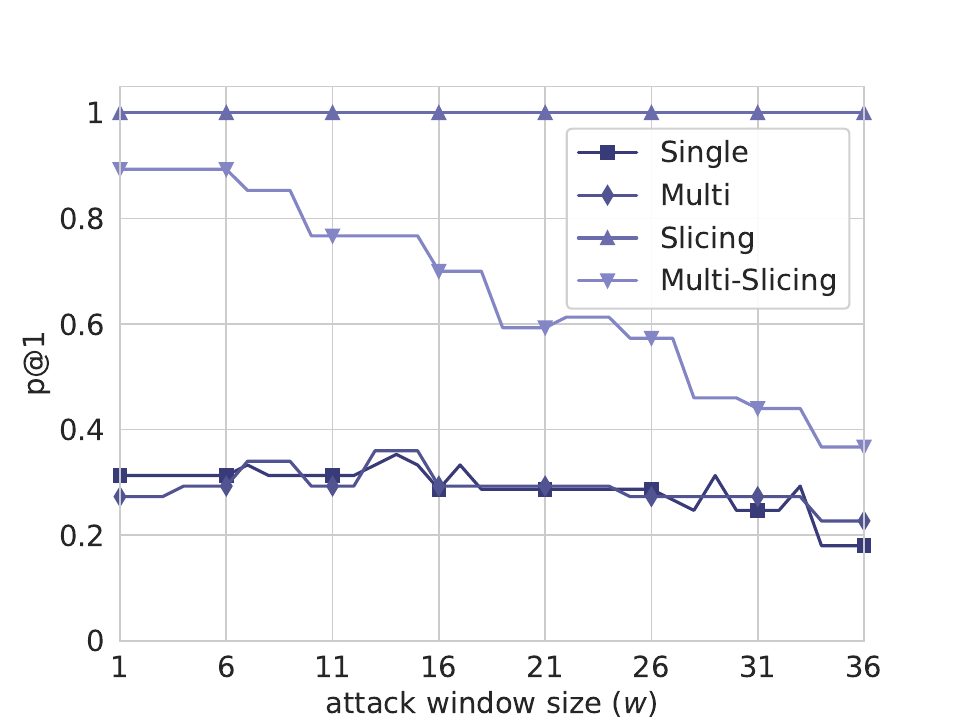}
        \caption{Evaluating attack window sizes $w$ regarding the p@1 scores of our attacks, used in \cref{sec:Attack Window Sizes}. The sizes 1--36 were tested for the Single- and the Slicing-Attack, while the sizes 1--12 were tested for both Multi-Attacks, which translate to the same total lengths due to their tripartite division of the attack data. We could however only evaluate every third step and therefore see plateaus. The attacks were performed on the WESAD dataset with a $DSF=1000$.}
        \label{app:fig:evaluation_attack_window_sizes}
        \Description{The Slicing-Attack is at 1.0 p@1 for all windows. The Multi-Attack and the Single attack hover around 0.3 but slightly decline with increased windows. The Multi-Slicing start high at 0.9 but falls quickly towards under 0.4 when attack window size rises.}
    \end{figure}

\begin{figure}[H]
        \centering
        \includegraphics[width=0.49\linewidth]{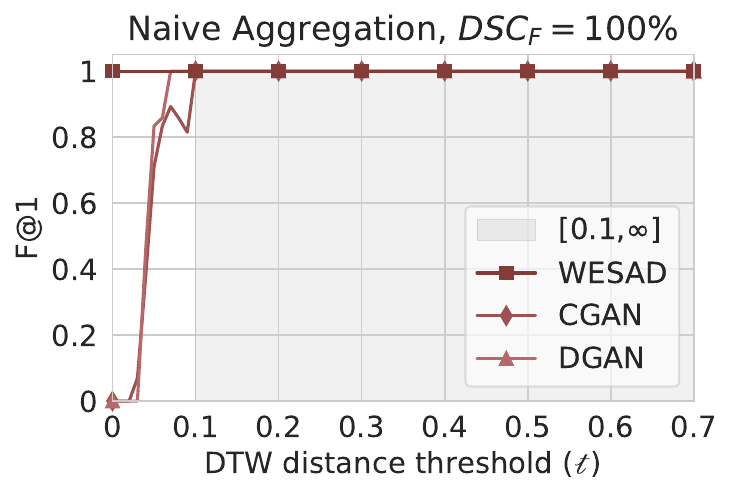} 
        \includegraphics[width=0.49\linewidth]{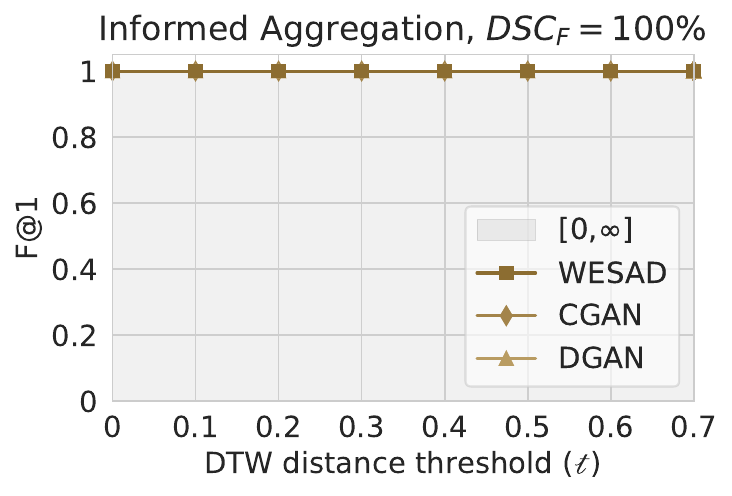}
        \caption{The "in-out" scenario on full overlap for \cref{sec:eval_in_out}. The x-axis gives the applied distance threshold and the y-axis shows the F@1 for our Slicing-DTW-Attack. We highlight the optimal threshold range across the tested datasets.}
        \label{app:fig:evaluation_in_out}
        \Description{Forn Naive WESAD is always at 1.0 F@1 across all thresholds. For The GANs we see a later but steep rise and both join WESAD in perfekt F@1 at a threshold of 0.1.}
    \end{figure}

In this appendix, we present two types of additional results that support our experiments.
First, for some of our methods we needed to tune hyperparameters and pick the corresponding best attack strategies for the framework from \cref{sec:Overview of the Attack Framework}.
These tuning results are given in \cref{app:tab:comparision_min_avg,app:tab:classes,app:tab:sensor_rank,app:fig:evaluation_attack_window_sizes}.
As a second type, we also offer further results that did not fit but are discussed in our results section in \cref{sec:evaluation}.
Those are found in \cref{app:tab:complexity,app:tab:attacks_eval,app:tab:evaluation_runtime,app:fig:evaluation_in_out}.

    \begin{table}[H]
        \centering
        \caption{Evaluation of the $\mathrm{avg}$ and $\mathrm{min}$ methods for Multi- and Slicing-DTW-Attacks as described in \cref{sec:dtw_attacks}.}
        \label{app:tab:comparision_min_avg}
        \begin{tabular}{c|ccc}
        \hline
            \textbf{Multi} & WESAD & CGAN$_{15}$ & DGAN$_{15}$ \\
          \hline
            average & \textbf{0.360} & \textbf{0.277} & \textbf{0.154} \\
            minimum & 0.360          & 0.219          & 0.133 \\
        \hline\hline
            \textbf{Slicing} & WESAD & CGAN$_{15}$ & DGAN$_{15}$ \\
          \hline
            average  & 0.953          & 1.000          & 1.000 \\
            minimum  & \textbf{1.000} & \textbf{1.000} & \textbf{1.000} \\
        \hline\hline
            \textbf{Multi-Slicing} & WESAD & CGAN$_{15}$ & DGAN$_{15}$ \\
          \hline
            average-average & 0.460          & 0.200          & 0.154 \\
            minimum-minimum & \textbf{0.893} & \textbf{0.296} & \textbf{0.343} \\
            average-minimum & 0.367          & 0.211          & 0.159 \\
            minimum-average & 0.500          & 0.200          & 0.154 \\
        \hline
        \end{tabular}
    \end{table}
    
    \begin{table*}[ht]
        \centering
        \caption{Results of our attacks using data from the non-stress or stress class as discussed in \cref{sec:Classes}. The table shows the p@k scores of the classes non-stress (non), stress (stress), and the weighted mean ($W_{class}$), which balances the class influence in our data. The attacks were performed on the WESAD dataset with a $DSF=1000$. Lower k-values are more important.}
        \label{app:tab:classes}
        \begin{tabular}{c|ccc|ccc|ccc|ccc}
        \hline
            \multirow{2}{*}{k} &
              \multicolumn{3}{c}{Single} &
              \multicolumn{3}{c}{Multi} &
              \multicolumn{3}{c}{Slicing} &
              \multicolumn{3}{c}{Multi-Slicing}\\
            & non & stress & $\mathcal{W}_{classes}$ & non & stress & $\mathcal{W}_{classes}$ & non & stress & $\mathcal{W}_{classes}$ & non & stress & $\mathcal{W}_{classes}$\\
          \hline
            k=1   & \textbf{0.294} & 0.276 & 0.289      & \textbf{0.339} & 0.178 & 0.291     & \textbf{1.000} & \textbf{1.000} & 1.000     & \textbf{0.767} & 0.411 & 0.660\\
            k=3   & 0.630 & 0.428 & 0.569               & 0.656 & 0.383 & 0.574              & 1.000 & 1.000 & 1.000                       & 0.967 & 0.661 & 0.875\\
            k=5   & 0.828 & 0.585 & 0.755               & 0.800 & 0.589 & 0.737              & 1.000 & 1.000  & 1.000                      & 0.994 & 0.800 & 0.936\\
        \hline
        \end{tabular}
    \end{table*}
    
    \begin{table*}[ht]
        \centering
        \caption{Sensor combinations for BVP (B), EDA (E), TEMP (T) and ACC (A), and p@k attack results using informed aggregation from \cref{sec:informed_sensor}. Attacks were performed on the WESAD dataset with a $DSF=1000$. Lower k-values are more important.}
        \label{app:tab:sensor_rank}
        \begin{tabular}{c|ccc|ccc|ccc|ccc}
        \hline
            \multirow{2}{*}{k} &
              \multicolumn{3}{c}{Single} &
              \multicolumn{3}{c}{Multi} &
              \multicolumn{3}{c}{Slicing} &
              \multicolumn{3}{c}{Multi-Slicing}\\
            & k=1 & k=3 & k=5 & k=1 & k=3 & k=5 & k=1 & k=3 & k=5 & k=1 & k=3 & k=5\\
          \hline
            B       & \textbf{1.000} & 1.000 & 1.000 & \textbf{1.000} & 1.000 & 1.000 & 0.997 & 1.000 & 1.000          & \textbf{0.991} & 1.000 & 1.000\\
            E       & 0.067          & 0.217 & 0.387 & 0.067          & 0.215 & 0.387 & 0.993 & 1.000 & 1.000          & 0.174          & 0.351 & 0.471\\
            T       & 0.084          & 0.250 & 0.433 & 0.104          & 0.233 & 0.383 & 0.996 & 1.000 & 1.000          & 0.045          & 0.172 & 0.307\\
            A       & 0.276          & 0.465 & 0.603 & 0.268          & 0.485 & 0.634 & \textbf{1.000} & 1.000 & 1.000 & 0.547          & 0.793 & 0.903\\
            B+E     & 0.176          & 0.539 & 0.757 & 0.176          & 0.555 & 0.757 & \textbf{1.000} & 1.000 & 1.000 & 0.701          & 0.894 & 0.947\\
            B+T     & 0.338          & 0.648 & 0.811 & 0.335          & 0.656 & 0.762 & \textbf{1.000} & 1.000 & 1.000 & 0.508          & 0.758 & 0.874\\
            B+A     & 0.609          & 0.887 & 0.963 & 0.637          & 0.878 & 0.958 & \textbf{1.000} & 1.000 & 1.000 & 0.923          & 0.975 & 0.987\\
            E+T     & 0.100          & 0.186 & 0.404 & 0.065          & 0.194 & 0.359 & \textbf{1.000} & 1.000 & 1.000 & 0.077          & 0.197 & 0.358\\
            E+A     & 0.104          & 0.259 & 0.522 & 0.094          & 0.260 & 0.482 & \textbf{1.000} & 1.000 & 1.000 & 0.415          & 0.692 & 0.783\\
            T+A     & 0.176          & 0.429 & 0.542 & 0.189          & 0.424 & 0.538 & \textbf{1.000} & 1.000 & 1.000 & 0.297          & 0.524 & 0.707\\
            B+E+T   & 0.260          & 0.540 & 0.706 & 0.233          & 0.528 & 0.698 & \textbf{1.000} & 1.000 & 1.000 & 0.506          & 0.799 & 0.902\\
            B+E+A   & 0.217          & 0.603 & 0.823 & 0.213          & 0.614 & 0.810 & \textbf{1.000} & 1.000 & 1.000 & 0.822          & 0.910 & 0.960\\
            B+T+A   & 0.411          & 0.690 & 0.807 & 0.440          & 0.655 & 0.770 & \textbf{1.000} & 1.000 & 1.000 & 0.715          & 0.873 & 0.920\\
            E+T+A   & 0.120          & 0.285 & 0.444 & 0.083          & 0.280 & 0.412 & \textbf{1.000} & 1.000 & 1.000 & 0.300          & 0.562 & 0.676\\
            B+E+T+A & 0.289          & 0.569 & 0.755 & 0.290          & 0.574 & 0.737 & \textbf{1.000} & 1.000 & 1.000 & 0.660          & 0.875 & 0.936\\
        \hline
        \end{tabular}
    \end{table*}
    
    \begin{table}[H]
        \centering
        \caption{Complexity reduction results for \cref{sec:evaluation_standard} regarding our datasets with 15 subjects, where we average the p@1 over our attack types. We test downsampling factors (DSF) and take the best ($DSF=1000$) for evaluating DBA and PCA.}
        \label{app:tab:complexity}
        \begin{tabular}{c|ccc}
        \hline
            DSF & WESAD & CGAN$_{15}$ & DGAN$_{15}$ \\
          \hline 
            1    & 0.551          & 0.372          & 0.340 \\
            10   & 0.593          & 0.440          & 0.391 \\
            100  & 0.625          & 0.436          & 0.372 \\
            1000 & \textbf{0.652} & \textbf{0.460} & \textbf{0.419} \\
          \hline 
            DBA  & 0.520          & 0.356          & 0.192 \\
            PCA  & 0.227          & 0.184          & 0.130 \\
        \hline
        \end{tabular}
    \end{table}
    
    \begin{table*}[ht]
        \centering
        \caption{For \cref{sec:evaluation_standard}, we gather the best attack sample window sizes $w$ and informed sensor combinations for BVP (B), EDA (E), TEMP (T) and ACC (A) for the evaluated datasets and attacks: Single (S), Multi (M), Slicing (SL), Multi-Slicing (MS). Window sizes for Multi-Attacks are given as their total over the three subsets.}
        \label{app:tab:attacks_eval}
        \begin{tabular}{c| c|c|c|c| c|c|c|c| c|c|c|c| c|c|c|c| c|c|c|c}
        \hline
            \multirow{2}{*}{} &
              \multicolumn{4}{c|}{WESAD} &
              \multicolumn{4}{c|}{CGAN$_{15}$} &
              \multicolumn{4}{c|}{DGAN$_{15}$} &
              \multicolumn{4}{c|}{CGAN$_{10000}$} &
              \multicolumn{4}{c}{DGAN$_{10000}$} \\
                & S & M & SL & MS    & S & M & SL & MS    & S & M & SL & MS    & S & M & SL & MS    & S & M & SL & MS \\
            \hline
            \textbf{size $w$} & 14 & 15 & 1 & 3    & 4 & 15 & 1 & 21     & 5 & 12 & 1 & 18     & 20 & 3 & 34 & 30     & 23 & 12 & 32 & 15   \\
          \hline\hline
            \textbf{Sensors} & \multicolumn{20}{c}{BVP (B), EDA (E), TEMP (T), ACC (A)} \\ \hline
            B       & \checkmark & \checkmark &  & \checkmark       &  &  &  & \checkmark                    & \checkmark &  &  &           &  &  &  &                   &  &  &  &  \\ \hline
            E       &  &  &  &                                      &  &  &  &                               &  &  &  &                     &  &  & \checkmark &         &  &  & \checkmark &  \\ \hline
            T       &  &  &  &                                      &  &  &  &                               &  &  &  &                     &  &  & \checkmark &                   &  &  & \checkmark &  \\ \hline
            A       &  &  & \checkmark &                            &  &  & \checkmark &                     &  &  & \checkmark &           &  &  & \checkmark &                   &  &  & \checkmark &  \\ \hline
            B+E     &  &  & \checkmark &                                      &  &  &  &                               &  &  &  &                     &  &  & \checkmark &                   &  & \checkmark & \checkmark &  \\ \hline
            B+T     &  &  & \checkmark &                                      &  &  &  &                               &  &  &  &                     &  &  & \checkmark &                   &  &  & \checkmark &  \\ \hline
            B+A     &  &  & \checkmark &                                      & \checkmark & \checkmark &  &           &  & \checkmark &  &           & \checkmark &  & \checkmark &         &  &  & \checkmark &  \\ \hline
            E+T     &  &  & \checkmark &                                      &  &  & \checkmark &                               &  &  & \checkmark &                     &  &  & \checkmark &                   &  &  & \checkmark &  \\ \hline
            E+A     &  &  & \checkmark &                                      &  &  & \checkmark &                               &  &  & \checkmark &                     &  &  & \checkmark &                   &  &  & \checkmark &  \\ \hline
            T+A     &  &  & \checkmark &                                      &  &  & \checkmark &                               &  &  & \checkmark &                     &  &  & \checkmark &                   &  &  & \checkmark &  \\ \hline
            B+E+T   &  &  & \checkmark &                                      &  &  &  &                               &  &  &  &                     &  &  & \checkmark & \checkmark        &  &  & \checkmark &  \\ \hline
            B+E+A   &  &  & \checkmark &                                      &  &  & \checkmark &                               &  &  &  &                     &  & \checkmark & \checkmark &         &  &  & \checkmark & \checkmark \\ \hline
            B+T+A   &  &  & \checkmark &                                      &  &  &  &                               &  &  &  &                     &  &  & \checkmark &                   & \checkmark &  & \checkmark &  \\ \hline
            E+T+A   &  &  & \checkmark &                                      &  &  & \checkmark &                               &  &  & \checkmark &                     &  &  & \checkmark &                   &  &  & \checkmark &  \\ \hline
            B+E+T+A &  &  & \checkmark &                                      &  &  & \checkmark &                               &  &  &  & \checkmark          &  &  & \checkmark &                   &  &  & \checkmark &  \\ 
        \hline
        \end{tabular}
    \end{table*}
    
    \begin{table*}[ht]
        \centering
        \caption{Runtime results in seconds for our attacks on datasets with different sizes, given by their subject counts, as evaluated in \cref{sec:evaluation_standard}. We give separate times for the attack and ranking process, as well as, their total. We used the external setting from \cref{sec:data model} assuming a single attack. We used the maximum window sizes $w$ for each attack from \cref{app:tab:attacks_eval} in this \cref{app:results}.}
        \label{app:tab:evaluation_runtime}
        \begin{tabular}{c|cccc|cccc|cccc|cccc}
        \hline
            \multirow{2}{*}{Scope} &
              \multicolumn{4}{c}{Single} &
              \multicolumn{4}{c}{Multi} &
              \multicolumn{4}{c}{Slicing} &
              \multicolumn{4}{c}{Multi-Slicing}\\
            & 15 & 100 & 1k & 10k & 15 & 100 & 1k & 10k & 15 & 100 & 1k & 10k & 15 & 100 & 1k & 10k \\
          \hline
            Attack  & 0.119 & 0.754 & 7.590 & 76.92 & 0.084 & 0.497 & 4.966 & 50.63 & 0.154 & 0.988 & 10.15 & 102.8 & 1.142 & 7.615 & 76.81 & 776.2 \\
            Ranking & 0.001 & 0.005 & 0.157 & 18.47 & 0.001 & 0.006 & 0.161 & 19.06 & 0.001 & 0.005 & 0.160 & 19.11 & 0.001 & 0.006 & 0.162 & 18.89 \\
            $\sum$   & 0.12 & 0.759 & 7.747 & 95.39 & 0.085 & 0.503 & 5.127 & 69.69 & 0.155 & 0.993 & 10.31 & 121.9 & 1.143 & 7.621 & 76.97 & 795.1 \\
        \hline
        \end{tabular}
    \end{table*}
    
\end{document}